\documentclass[aps,twocolumn,pra,showpacs,superscriptaddress,floatfix]{revtex4-2}
\usepackage{amssymb,amsmath,amsfonts}
\usepackage{epsfig,graphicx}
\usepackage{hyperref}
\usepackage[dvipsnames]{xcolor}

\hypersetup{colorlinks,citecolor=NavyBlue,filecolor=black,linkcolor=NavyBlue,urlcolor=NavyBlue}

\newcommand\ha{\hat{a}}
\newcommand\hadag{\hat{a}^{\dagger}}
\newcommand\hadaga{\hat{a}^{\dagger} \hat{a}}

\newcommand\hb{\hat{b}}
\newcommand\hbdag{\hat{b}^{\dagger}}
\newcommand\hn{\hat{n}}
\newcommand\ket[1]{\left|\textstyle{#1}\right\rangle}
\newcommand\bra[1]{\left\langle\textstyle{#1}\right|}

\newcommand\hx{\hat{x}}
\newcommand\hp{\hat{p}}

\usepackage{soul}
\usepackage{color}
\definecolor{blueRP}{rgb}{0.0, 0.58, 0.71}
\definecolor{blueCh}{rgb}{0.2, 0.4, 0.7}

\newcommand{\changes}[1]{\textcolor{black}{#1}}

\begin{document}
\title{Quantum metrological advantage of high-order squeezed states}

\author{Rub{\'e}n Gordillo-Hachuel}
\email{rgordill@ing.uc3m.es}
\affiliation{Department of Physics, Universidad Carlos III de Madrid, Avda. de la Universidad 30, Legan\'es, 28911  Madrid, Spain}
\affiliation{Department of Electronic Technology, Universidad Carlos III de Madrid, Avda. de la Universidad 30, Legan\'es, 28911  Madrid, Spain}
\author{Erik Torrontegui}
\affiliation{Department of Physics, Universidad Carlos III de Madrid, Avda. de la Universidad 30, Legan\'es, 28911  Madrid, Spain}
\author{Cristina de Dios}
\affiliation{Department of Electronic Technology, Universidad Carlos III de Madrid, Avda. de la Universidad 30, Legan\'es, 28911  Madrid, Spain}
\author{Ricardo Puebla}
\email{rpuebla@fis.uc3m.es}
\affiliation{Department of Physics, Universidad Carlos III de Madrid, Avda. de la Universidad 30, Legan\'es, 28911  Madrid, Spain}

\begin{abstract}
Quantum correlations can be harnessed to improve the precision in parameter estimation beyond classical capabilities. Under a standard interferometric or rotation protocol, it is well established that the optimal single-mode Gaussian state is a standard squeezed vacuum, which enables Heisenberg limited precision. In this work, we investigate the potential metrological advantage of two distinct families involving high-order squeezing, namely, \textit{m}th-phase and multisqueezed states. Our results show that these non-Gaussian states can grant  a significant metrological advantage with respect to the optimal squeezed vacuum under equivalent conditions, i.e. at equal occupations. Their advantage holds both at low and large occupations, but its behavior critically depends on the chosen family of high-order squeezing. While higher squeezing orders enhance the advantage, this comes  at the cost of higher-order observables in the measurement for full metrological performance. Finally, we study their robustness to standard decoherence channels, i.e. pure dephasing and zero-temperature damping. Employing standard squeezing as reference state, our results indicate a reasonable robustness against damping up to a certain noise strength, while their metrological advantage becomes fragile under pure dephasing. Our work shows the potential enhancement in quantum metrology beyond Gaussian states, carefully detailing the main challenges and limitations. 
\end{abstract}
\maketitle
\section{Introduction}

    Quantum metrology is one of the most active research areas within quantum technologies~\cite{Giovannetti2004,Giovannetti2006,Paris2009,Giovannetti2011,Degen2017,Pezze2018}. The main goal consists in devising protocols to improve the precision in parameter estimation to overcome the so-called standard quantum limit (SQL). Indeed, by harnessing quantum resources, such as entanglement, it has been shown that the SQL can be beaten~\cite{Huelga1997,Giovannetti2004}, leading to the Heisenberg limit for precision. A quantum advantage is achieved in this scenario since the variance of the estimated parameter scales as $1/N^2$ (Heisenberg limit) rather than as $1/N$ for the SQL, where $N$ refers to the number of probes. Depending on how the target parameter is encoded in the quantum state and on the probe interactions, the precision may be enhanced even further~\cite{Roy2008}. 

    The quantum Fisher information (QFI) plays a pivotal role when characterizing the metrological usefulness of a particular protocol. Indeed, as dictated by the celebrated Cram\'er-Rao bound, the inverse of the QFI sets the ultimate precision limit in parameter estimation. In this manner, different strategies have been proposed to boost the QFI under equivalent number of resources, such as number of probes. One of the standard metrological protocols consists in preparing a state $\hat{\rho}$, which then undergoes a time evolution where the target parameter is the accumulated phase $\theta$. This evolution imprints a dependence on the target parameter,  $\hat{\rho}_\theta$. Depending on the nature of the state $\hat{\rho}$, the precision to estimate $\theta$ can be increased with respect to a classical scenario. This standard metrological scheme is commonly known as interferometric or rotation protocol.  

    The number $N$ of different spin-like quantum register is a good quantity to analyze the dependence of the QFI. Indeed, certain entangled states among $N$ qubits enable to saturate the Heisenberg limit for precision, allowing even to define optimal states for quantum metrology~\cite{Fiderer2019}. In continuous-variable systems, however, a quantification of the available metrological resources must focus on the average occupation, which is typically related to\changes{---and limited by---}their energy~\cite{Gorecki2025, Fadel_2025, Sahota_2015, Perola2026}.  In this manner, \changes{the standard is to compute} the QFI in terms of the average number of bosonic excitations $\langle \hn \rangle$, \changes{establishing} an analogy with the number of registers $N$ in discrete quantum systems. Hence, the key figure of merit consists in maximizing the QFI for a fixed occupation number $\langle \hn \rangle$. Among all possible states, those with a Gaussian Wigner function are of particular relevance~\cite{Ferraro}, such as standard coherent and squeezed states. Indeed, the squeezed vacuum is known to be the optimal Gaussian state for metrological tasks in a rotation protocol~\cite{Pinel2013,Matsubara2019}. By harnessing quantum correlations among Fock states, such squeezed states feature a QFI reaching the Heisenberg limit as their average occupation increases. The natural question is therefore whether departing from Gaussian states can grant any meaningful metrological advantage over squeezed vacuum states, and if so, for what type of non-Gaussian families. It is worth mentioning that, although states featuring large fluctuations in the occupation number can surpass the performance of squeezed vacuum states~\cite{Shapiro1989}, they may remain ineffective under specific measurement schemes~\cite{Braunstein1992}.

    In this article we tackle this precise question, analyzing two distinct families of non-Gaussian states. In particular, we focus on two nonequivalent generalizations of coherent and squeezed states to higher-order bosonic operators, namely \textit{m}th-phase and multisqueezed states. Such non-Gaussian states display a large amount of Wigner negativity. 
    The interest in these states began in the '80s with a discussion about their existence and properties~\cite{Fisher1984, Braunstein1987, Wu1986, Hillery1990,Banaszek1997}. Recently, they have attracted considerable attention ~\cite{McConnell2022,3ps_Gessner_2025,Ashhab_2025,Gordillo2026,Ashhab2026}, since they can be generated in spin-boson systems~\cite{Chang2020, Gasparinetti_2022, Eriksson_2024, Bazavan2024, saner2024}. In particular, one of the states within the \textit{m}th-phase family, the cubic-phase state plays a key role in quantum computation with continuous-variables systems as it unlocks universal computation~\cite{Gottesman2001,Gu2009,Zheng2021}. Here we focus on the potential metrological advantage, and as we show,  both families can improve the optimal Gaussian QFI,  i.e. that of squeezed vacuum states. The \textit{m}th-phase family presents a better QFI than the squeezed vacuum at any occupation, leading to a Heisenberg limited precision. Yet, multisqueezed states of order $m>2$ demand a careful numerical study to ensure numerical convergence, requiring to explicitly include a pump  field that enables the corresponding $m$-photon down conversion~\cite{Hillery1990}. Under these conditions, we find that the QFI also surpasses that of  the squeezed vacuum for reasonably low occupations, displaying a remarkably different behavior with respect to \textit{m}th-phase states. A further study of the sensitivity under a restricted set of measurements indicates a useful advantage,  at the expense of measuring complex combination of observables. Finally, we also analyze the robustness under two relevant decoherence mechanisms, namely, dephasing and zero-temperature damping. As expected, the metrological advantage for both families of high-order squeezing is fragile under dephasing, while it is shown to be more robust against zero-temperature damping.

    The article is organized as follows. First, in Sec.~\ref{s:int}, we introduce the standard interferometric protocol and the quantum Fisher information. In Sec.~\ref{s:NGstates}, the \textit{m}th-phase and multisqueezed states are defined, studying their maximum metrological capabilities, emphasizing the improvement with respect to squeezed vacuum states at a fixed average occupation. The useful metrological advantage of such families of non-Gaussian states is discussed in Sec.~\ref{s:adv}, where we analyze the sensitivity for measurements resulting from an optimal linear combination over a set of observables. In Sec.~\ref{s:decoh} we show the results of two relevant decoherence channels. Finally, the main results and conclusions of this work are presented in Sec.~\ref{s:conc}.

\section{Interferometric protocol and quantum Fisher information}\label{s:int}


    As mentioned above, we consider a standard rotation or interferometric metrological protocol. Here we briefly review this theoretical framework, introducing the QFI. In this standard protocol, a single-bosonic mode is initialized in a state $\hat{\rho}$, described in terms of the usual annihilation and creation operators $\ha$ and $\hadag$, fulfilling $[\ha,\hadag]=1$. Then, the target parameter $\theta$ is encoded as a phase shift or rotation of the state $\hat{\rho}$, so that
    \begin{align}
        \hat{\rho}_\theta=\hat{U}_\theta \hat{\rho} \hat{U}_\theta^\dagger,
    \end{align}
    where $\hat{U}_\theta=e^{-i\theta \hat{H}}$ being $\hat{H}$ the generator of the rotation. In this manner, the parameter $\theta$ is encoded in the quantum state of the system. The sensitivity of the state, or more precisely, its metrological usefulness is intimately related with the distinguishability between the states $\hat{\rho}_\theta$ and $\hat{\rho}_{\theta+\delta_\theta}$. Therefore, the sensitivity is directly related to the QFI, which according to the Cram\'er-Rao bound quantifies the best achievable precision in parameter estimation~\cite{Braunstein94,Giovannetti2004,Giovannetti2006,Paris2009}. The QFI with respect to the target parameter $\theta$ can then be written in general as~\cite{Braunstein94,Taddei13}
    \begin{align}
        F_Q=\left.-4\frac{\partial^2 \mathcal{F}(\hat{\rho}_\theta,\hat{\rho}_{\theta+\delta_\theta})}{\partial \ \delta_\theta^2}\right|_{\delta_\theta=0},
    \end{align}
    in terms of the state fidelity $\mathcal{F}(\hat{\rho},\hat{\sigma})={\rm Tr}[\sqrt{\sqrt{\hat{\rho}}\hat{\sigma}\sqrt{\hat{\rho}}}]$.  This rotation protocol encodes the target parameter as a unitary transformation, thus diagonalizing the state as $\hat{\rho}=\sum_{k=1}^d p_{k} \ket{\psi_k}\bra{\psi_k}$, being $d$  the dimension of the Hilbert space, this expression simplifies to 
    \begin{align}\label{eq:QFI_mix}
        F_Q=2\sum_{k,l}^d p_{k,l}|\bra{\psi_k}\hat{H}\ket{\psi_l}|^2, 
    \end{align}
    with 
    \begin{align}
        p_{k,l}=\begin{cases}
        0 \quad &{\rm if} \quad p_k=p_l=0,\\
        \frac{(p_k-p_l)^2}{p_k+p_l} \quad &{\rm otherwise}.
        \end{cases}
    \end{align}
    For pure states, $\hat{\rho} = \ket{\psi} \bra{\psi}$, $F_Q$ further simplifies  to
    \begin{align}\label{eq:QFI_general}
        F_Q=4\left( \bra{\psi}\hat{H}^2\ket{\psi}-|\bra{\psi}\hat{H}\ket{\psi}|^2\right),
    \end{align}
    that is, the QFI is simply proportional to the variance of the operator $\hat{H}$ over the initial state $\ket{\psi}$.

    To better quantify the role of the QFI one must carefully analyze  its behavior in terms of the employed resources~\cite{Giovannetti2006}. In discrete or spin-like systems, the number of probes or registers $N$ appears as the key resource. This allows to identify quantum-enhanced metrological regimes, that is, scenarios that can surpass the classical limit $F_Q\propto N$ (SQL), to reach a Heisenberg-limited QFI, $F_Q\propto N^2$~\cite{Giovannetti2006}.  Although this notion is blurred  in bosonic systems, the average occupation number $\langle \hat{n}\rangle\equiv \langle \hadaga \rangle$ plays here the role of the resources or number of probes in discrete systems. 
    In this manner, the key figure of merit consists in maximizing the QFI for a fixed  $\langle \hn \rangle$.

    Under this rotation protocol, considering $\hat{H} = \hadaga$, one recovers the well-known results for single-mode Gaussian states, i.e. coherent and standard squeezed states (see App.~\ref{app:a}). Coherent states $\ket{\alpha}=e^{\alpha \hat{a}^\dagger-\alpha^\star \hat{a}}\ket{0}$ feature a QFI given by 
    \begin{align}\label{eq:FQC}
        F_Q^C=4\langle \hn \rangle,
    \end{align}
    while the corresponding QFI for a squeezed vacuum $\ket{\zeta}=e^{\frac{1}{2}(\zeta^\star \hat{a}^2-\zeta \hat{a}^{\dagger,2})}\ket{0}$ reads as
    \begin{align}\label{eq:FQS}
        F_Q^S=8(\langle \hn \rangle^2+\langle \hn \rangle).
    \end{align}
    In the following, we will refer to the squeezed vacuum as the standard squeezed state, in order to stress the difference with respect to the other families of generalized and high-order squeezing. 
    The upper-script $C$ and $S$ denotes coherent and standard squeezed state, respectively. 
    For $\langle \hn \rangle\gg 1$ the standard squeezed states provide a Heisenberg-limited precision ($F_Q^S\propto \langle \hn \rangle^2$), while coherent states saturate at the SQL ($F_Q^C\propto \langle \hn \rangle$), being standard squeezed states the optimal Gaussian states for quantum metrology in a rotation protocol~\cite{Pinel2013,Matsubara2019}. 

\section{Metrology with Non-Gaussian States}\label{s:NGstates}

    Since standard squeezed states are optimal when restricting to Gaussian states, this raises the natural question on to what extent this metrological enhancement can be improved when departing from this restrictive set, i.e. considering non-Gaussian states. Such potential improvement must be analyzed under equal resources, which in this case denotes a fixed occupation $\langle \hat{n} \rangle$.

    Non-Gaussian states are characterized for having negativity regimes in their associated Wigner function~\cite{Banaszek1997}. This negativity has been recognized as a useful resource~\cite{Albarelli2018,Takagi2018}, which may imply greater distinguishability between nearby states, potentially leading to an improvement in the QFI. However, it is important to remark that there is no direct relation between Wigner negativity and usefulness in a rotation protocol. For example, Fock states $\ket{n>1}$ are known to exhibit a large Wigner negativity~\cite{Albarelli2018}, yet, they are poor states for this metrological protocol. Indeed, $\ket{n}$ leads to $F_Q=0$ since $\langle \hat{n}^2\rangle=\langle \hn \rangle^2=n^2$. \changes{Therefore, the relation between the metrological advantage and Wigner negativity strongly depends on the employed protocol and the considered state.} In this article we investigate the metrological performance of two distinct families of non-Gaussian states, namely, \textit{m}th-phase and multisqueezed states, \changes{under the rotation protocol. However,} we remark that other types of non-Gaussian states may be useful under different metrological protocols~\cite{Guo2024,Rahman2025}. These families refer to two unequal generalizations of coherent and standard squeezed states to high-order squeezing, which have attracted considerable attention over the years~\cite{Fisher1984,Braunstein1987,Wu1986,Hillery1990,Banaszek1997,McConnell2022,Bazavan2024,3ps_Gessner_2025,Ashhab_2025,Gordillo2026,Ashhab2026}. Although these families are frequently grouped together, they exhibit substantially different behaviors, as we discuss in detail in the following.

    \subsection{${\bf m}$th-phase states}

        The family of \textit{m}th-phase states refers to a possible generalization of coherent and standard squeezed states to  higher orders of $\hadag$ and $\ha$, by increasing the order of the full operator $(\hadag + \ha)$. This family is then generated by the operator
        \begin{align}\label{eq:Umphase}
            \hat{U}_m(\gamma_m) = \exp{\Bigg\{\frac{i\gamma_m}{2^{m/2}} (\ha +  \ha^{\dagger})^m\Bigg\}},
        \end{align}
        where $m=0,1,2,3\ldots$ is the squeezing order, and $\gamma_m$ the corresponding generalized squeezing strength or parameter. Although in general $\gamma_m\in\mathbb{C}$, for the analysis carried out in this article, one can restrict without loss of generality to non-negative real values, $\gamma_m\geq 0$. 

        In particular, we refer to \textit{m}th-phase states as the operator in Eq.~\eqref{eq:Umphase} applied to the vacuum, that is
        \begin{align}\label{eq:gammam}
            \ket{\gamma_m}=\hat{U}(\gamma_m)\ket{0}=\exp{\Bigg\{\frac{i\gamma_m}{2^{m/2}} (\ha +  \ha^{\dagger})^m\Bigg\}}\ket{0}.
        \end{align}
        Besides the trivial $m=0$ case, $m=1$ simply leads to a coherent state, while it becomes a standard squeezed state for $m=2$, both with a rotation with respect to their usual expressions. That is, the states $\ket{\gamma_{1,2}}$ lead to the well-known QFI under the interferometric protocol, given in Eqs.~\eqref{eq:FQC} and~\eqref{eq:FQS}, respectively. Therefore, we will refer to \textit{m}th-phase states as those with $m\geq 3$, and in particular, we will  focus on cubic-phase $(m=3)$ and quartic-phase $(m=4)$ states.  
        
        In general, it is possible to analytically compute the QFI for any \textit{m}th-phase state. For that, we shift to the position representation, that allows us to calculate the average of the photon occupation $\hat{n} = \hadaga$ and its square $\hat{n}^2 = (\hadaga)^2$. We refer  to App.~\ref{app:mth-phase} for the details of the calculations. These expectation values for $m\geq 3$ result in
        \begin{align}
            \langle \hn \rangle = m^2 \gamma_m^2 &\frac{(2m-3)!!}{2^m},\\
            \langle \hat{n}^2\rangle = m^4 \gamma_m^4 &\frac{(4m-5)!!}{2^{2m}} \nonumber\\&+ m^2 \gamma_m^2 \frac{(2m-5)!!}{2^m} (m^2-2).
        \end{align}
        Note that $k!!=k(k-2)(k-4)\dots$ denotes the double factorial operation of $k$. Then, from the previous results and using Eq.~\eqref{eq:QFI_general}, it is straightforward to obtain the QFI for \textit{m}th-phase states in terms of the occupation $\langle \hn \rangle$, which we denote as $F_Q^{mp}$, i.e.
        \begin{align}\label{eq:FQm}
            F_Q^{mp} = 4\left(\frac{ (4m-5)!!}{((2m-3)!!)^2}-1\right) \langle \hn \rangle^2 + \frac{4m^2-8}{2m-3} \langle \hn \rangle .
        \end{align}
         Although the calculation was made considering only $m \ge 3$, the resulting QFI is also valid for the case $m = 2$, as it can be verified comparing this result with $F_Q^S=F_Q^{2p}$, given in Eq.~\eqref{eq:FQS}. Hence, for a low occupation regime, \textit{m}th-phase states display a SQL behavior, that is $F_Q^{mp} \propto \langle \hn \rangle$, while in the high occupation regime, they exhibit a Heisenberg-limited QFI,  $F_Q^{mp} \propto \langle \hn \rangle^2$. 
        
        Furthermore, for any non-zero occupation, the QFI of a \textit{m}th-phase state increases exponentially with $m$,  $F_Q^{mp}\propto 4^m$ for $m\gg 1$. This can be shown as follows. For any non-zero occupation $\langle \hat{n}\rangle>0$, the term $(4m-5)!!/((2m-3)!!)^2$ in Eq.~\eqref{eq:FQm} will dominate for sufficiently large $m$. Employing the identity $(2n-1)!!=(2n)!2^{-n}/n!$, it follows that, at leading order $(4m-5)!!/((2m-3)!!)^2\sim (4m)! (m!)^2/((2m)!)^3\sim 4^m$, where the last step can be worked out employing the Stirling approximation $n!\sim \sqrt{2n\pi}(n/e)^n$. That is, for any non-zero occupation, $\langle \hn\rangle>0$, the \textit{m}th-phase QFI scales as $F_Q^{mp}\propto 4^m$ for a growing $m$. Therefore, the precision limit with which the parameter $\theta$ can be determined using \textit{m}th-phase states increases with $m$, being able to obtain any arbitrary improvement with respect to the standard squeezed states case at any given occupation $\langle \hat{n}\rangle$. \changes{This can be understood intuitively: as $m$ increases, the vacuum state $\ket{0}$ is deformed more strongly, thereby enhancing the distinguishability of the resulting states under rotations.} Already this family of non-Gaussian states reveals the absence of an upper limit for the QFI at any finite occupation. Although relevant, this remains as a theoretical curiosity since the generation of such states becomes difficult for growing $m$. In addition,  even if we assume the preparation of such states to be feasible, harnessing its full metrological advantage requires non-trivial measurements, increasing again in complexity with the order $m$. As mentioned above, we will focus on the first non-trivial cases that are more experimentally relevant, i.e. cubic- ($m=3$) and quartic-phase states ($m=4$), and discuss this latter fact in Sec.~\ref{s:mphase_sensitivity}. 

\begin{figure}
            \centering
            \includegraphics[width=\linewidth]{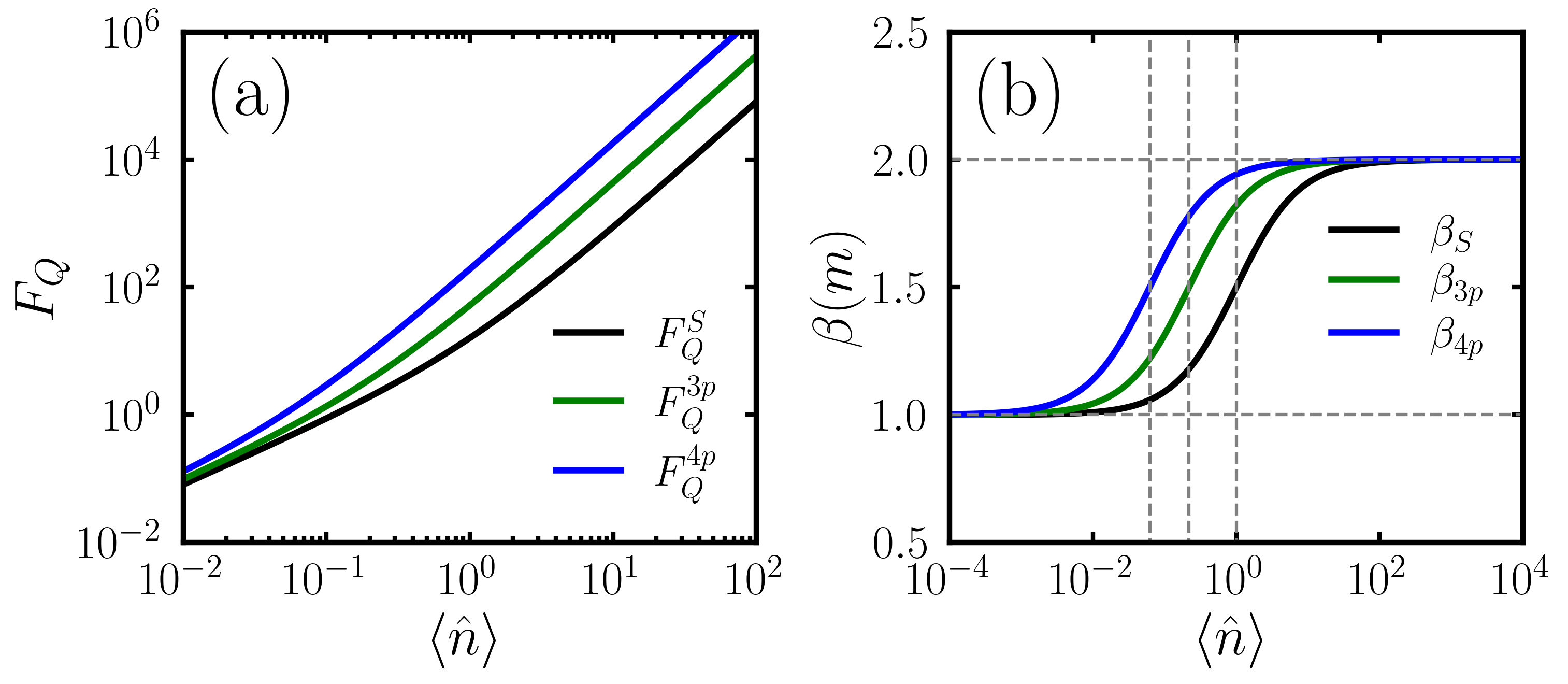}
            \caption{(a) QFI $F_Q^{mp}$ with respect to target parameter $\theta$, for cubic- ($F_Q^{3p}$, green line) and quartic-phase ($F_Q^{4p}$, blue line) states as a function of the occupation number $\langle \hat{n} \rangle$. For comparison, the solid black line shows the QFI for the standard squeezed state. (b) Scaling exponent $\beta(m)$ of the QFI, i.e. $F_Q^{mp}\propto \langle \hat{n}\rangle^{\beta(m)}$, for the states in (a) as a function of the occupation. Note how $F_Q^{mp}$ grows with increasing $m$ at a fixed occupation, while the Heisenberg limit region, $\beta(m)=2$, takes place at smaller occupations.  }
            \label{fig:QFI_mphase}
        \end{figure}

        For comparison, Fig.~\ref{fig:QFI_mphase}(a) shows the QFI for cubic- and quartic-phase states as a function of the average occupation $\langle \hn\rangle$, together with the QFI of the standard squeezed states. Clearly, at a fixed occupation, increasing $m$ leads to a larger QFI. In addition, the Heisenberg-limited region occurs at smaller occupations for growing $m$. This can be better seen in  Fig.~\ref{fig:QFI_mphase}(b), that shows the exponent $\beta(m)$ of $F_Q^{mp}$ at each occupation for a \textit{m}th-phase state, i.e. $F_Q^{mp}\propto \langle \hat{n}\rangle^{\beta(m)}$, which  reveals the transition from SQL ($\beta=1$) to the Heisenberg limit ($\beta=2$).  To quantify this transition, we compute the exponent as $\beta(m) =\frac{d\log_{10} F_Q^{mp}}{d\log_{10} \langle \hat{n} \rangle}$. Defining $F_Q^{mp}=c_2(m)\langle \hat{n}\rangle^2+c_1(m)\langle \hat{n}\rangle$, so that $c_{1,2}(m)$ denote the coefficients dependent on $m$ in Eq.~\eqref{eq:FQm}, one simply finds 
        \begin{align}\label{eq:beta}
            \beta(m)=2-\frac{c_1(m)}{c_2(m)\langle \hat{n}\rangle+c_1(m)}.
        \end{align} 
        Clearly, $1\leq \beta(m)\leq 2$. These states reach the Heisenberg limit, $\beta(m)=2$, at smaller occupations the larger $m$. Considering that the transition from SQL to Heisenberg limit takes places at the midpoint $\beta(m)=3/2$, we find that the occupation at such crossover decreases exponentially with $m$. For $m=2$, i.e. for the standard squeezed state, this takes place at $\langle \hat{n}\rangle=1$, while for cubic- and quartic-phase states at $\langle \hat{n}\rangle=7/32$ and $7/113$, respectively (cf. Fig.~\ref{fig:QFI_mphase}(b)). Doing a similar analysis as for the asymptotic QFI scaling, but using Eq.~\eqref{eq:beta}, we find that the transition from SQL to Heisenberg limit happens at occupations $\langle \hat{n}\rangle\sim 4^{-m}$ for $m\gg 1$.

    \subsection{Multisqueezed states}\label{s:multi_qfi}

        Multisqueezed states are yet another relevant generalization of coherent and standard squeezed states to higher-order terms. These states coincide with the standard Gaussian states at first and second order, but result in a nonequivalent family with respect to \textit{m}th-phase states for higher-than quadratic squeezing~\cite{Fisher1984,Banaszek1997,Hillery1990}. This family  describes a \textit{m}-photon down conversion process, whose generator can be written as
        \begin{align}
            \hat{U}_m(\xi_m) = e^{\xi_m^* \ha^m - \xi_m \ha^{\dagger, m}}.
        \end{align}
        As before, $m=0,1,2,3\ldots$ is the squeezing order, and $\xi_m$ corresponds to the high-order squeezing parameter. Similar to \textit{m}th-phase states,  one can restrict without loss of generality to non-negative real values, $\xi_m\geq 0$. In this manner, one may define the \textit{m}th-order multisqueezed state as
        \begin{align}\label{eq:xim}
            \ket{\xi_m}=e^{\xi_m(\ha^m - \ha^{\dagger,m})}\ket{0}.
        \end{align}
        Clearly, for $m=1$ we recover a coherent state, while for $m=2$ it becomes a standard squeezed  state. Unfortunately, and contrary  to \textit{m}th-phase states, for $m\geq 3$ there is no closed-form solution to the amplitudes over the Fock states and to their quadratures. Therefore, these states must be studied numerically, considering a truncated Hilbert space. Moreover, as shown in \cite{Hillery1990,Gordillo2026}, these states $\ket{\xi_{m\geq 3}}$ feature a diverging occupation at a finite value of $\xi_{m\geq 3}$, critically dependent on Fock-basis truncation. For this reason, as discussed in~\cite{Hillery1990}, one must explicitly include \changes{the} pump field \changes{that produces the \textit{m}-photon down conversion on the local oscillator, thus leading to high-order multisqueezed states. This will guarantee physically meaningful results and convergence of these states.} Considering a coherent state for the pump field, we ensure physically sound  \textit{m}th-order squeezed states, defined in the joint Hilbert space of local oscillator and pump $\mathcal{H}_{LO}\otimes \mathcal{H}_{P}$, i.e. 
        \begin{align} \label{eq:multi_pump}
            \ket{\kappa_m}=e^{\kappa_m(\ha^m \hbdag - \ha^{\dagger,m} \hb)} \ket{0, \alpha},
        \end{align}
        where $\kappa_m \ge 0$  is related to the high-order multisqueezed parameter given before $\xi_m$, and involves the coupling strength between the pump and local oscillator. Note that $\ha$ and $\hb$ are the annihilation operators corresponding to the local oscillator and pump field, respectively, while $\ket{\alpha} = \exp{\{ \alpha (\hbdag - \hb)\}}\ket{0}$ is the initial coherent state of the pump. \changes{This initial amplitude $\alpha$, is another metrological parameter related with $\langle \hn \rangle$, and with the generation of the multisqueezed state.} Without loss of generality, we take $\alpha\in \mathbb{R}$. It is important to mention that, although multisqueezed must be studied taking explicitly into account the pump to avoid potential divergences and numerical artifacts arising from a finite Fock-basis truncation, we focus only on the properties of the local oscillator. That is, the pump enables a $m$-photon down conversion (cf. Eq.~\eqref{eq:multi_pump}), but is then traced out to analyze the metrological capacity of the resulting state of the local oscillator, $\rho(\kappa_m)={\rm Tr}_{P}[\ket{\kappa_m}\bra{\kappa_m}]$, where ${\rm Tr}_P[\cdot]$ denotes the  partial trace over the pump Hilbert space.

\begin{figure}
            \centering
            \includegraphics[width=\linewidth]{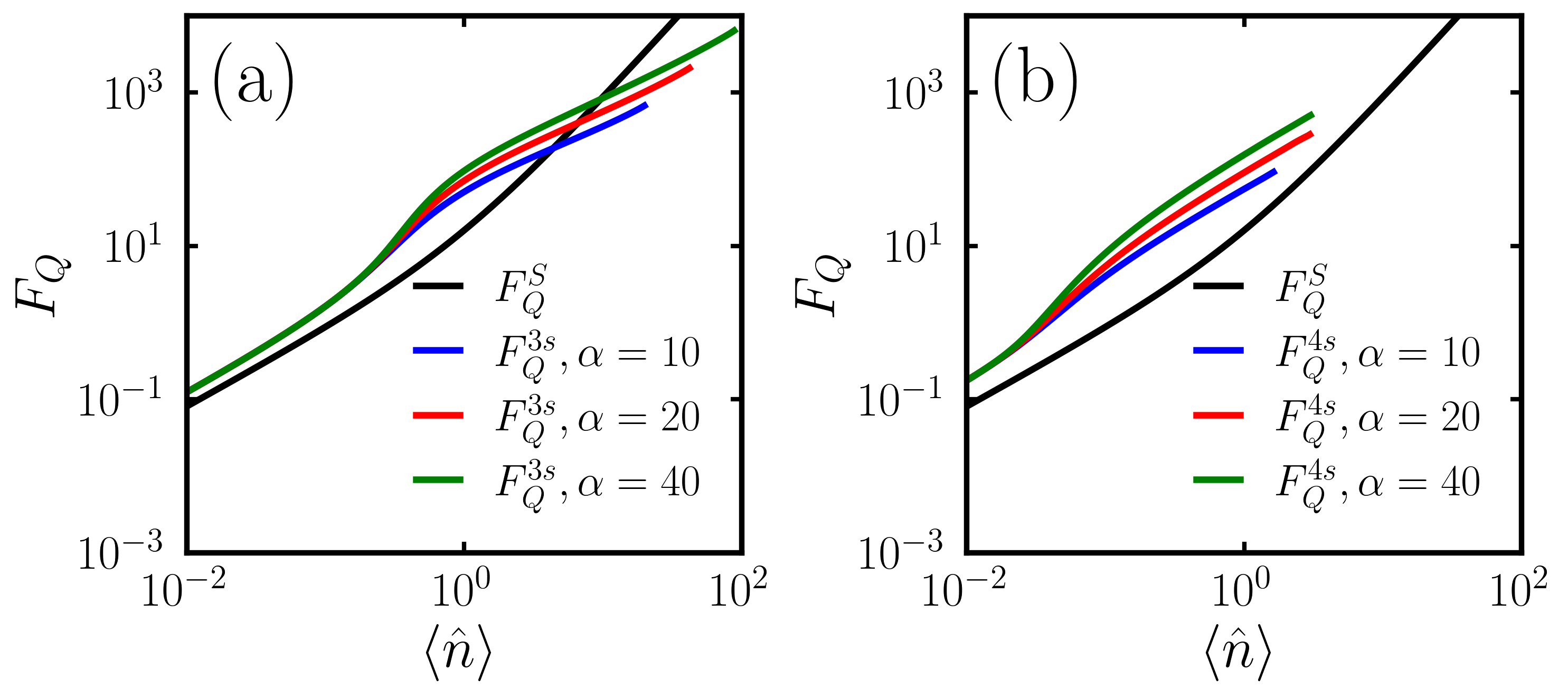}
            \caption{QFI of multisqueezed states $F_Q^{ms}$ under a rotation protocol with different initial pump amplitudes of a coherent state $\ket{\alpha}$ and as a function of the occupation number $\langle \hat{n} \rangle$. The solid black line shows the QFI for the standard squeezed state for comparison. Panel (a) shows the QFI of trisqueezed states $F_Q^{3s}$, while (b) corresponds to quartsqueezed states $F_Q^{4s}$. The lines for $F_Q^{ms}$ terminate at a finite occupation since a finite $\alpha$ restricts the maximum population for the generated multisqueezed state upon a $m$-photon down conversion. }
            \label{fig:QFI_multisqueezed}
        \end{figure}

        In the following we will focus on trisqueezed ($m=3$) and quartsqueezed ($m=4$) states, which  are numerically simulated employing Eq.~\eqref{eq:multi_pump}, always ensuring numerical convergence. The population over large Fock states is kept below or equal to $10^{-17}$ in both Hilbert spaces. In particular, for the local oscillator, we truncate the Fock basis up to $\ket{N}$, such that  $\bra{N}\rho(\kappa_m)\ket{N}\leq 10^{-17}$ being $N$ a multiple of $m$. Upon the $m$-photon down conversion, and tracing out the pump, the resulting mixed state for the local oscillator $\rho(\kappa_m)$ is then used in the rotation protocol. Clearly, this  state depends on the photons provided by the pump, i.e. on $\alpha$. The corresponding QFI of $\rho(\kappa_m)$, obtained via Eq.~\eqref{eq:QFI_mix}, is shown in Fig.~\ref{fig:QFI_multisqueezed}(a) and (b) for trisqueezed and quartsqueezed states, respectively, for different initial coherent states $\ket{\alpha}$ of the pump field. In order to make clear the distinct high-order squeezing family, the QFI of multisqueezed states is written as $F_{Q}^{ms}$, in contrast to $F_{Q}^{mp}$ that refers to \textit{m}th-phase states.

        The behavior of the QFI for trisqueezed states, $F_{Q}^{3s}$ is shown in Fig.~\ref{fig:QFI_multisqueezed}(a)) as a function of the average occupation $\langle \hat{n}\rangle$. For the considered pump field, $\alpha=10$, $20$ and $40$, at low occupations, $\langle \hn \rangle\lesssim 10^{-1}$, trisqueezed states achieve a SQL ($F_Q^{3s} \propto \langle \hn \rangle$), just as the standard squeezed state $F_Q^S$ but with a constant improvement, i.e. in this region $F_Q^{3s}/F_Q^S\approx 1.5$. Now, for states with occupation $10^{-1}\lesssim \langle \hat{n}\rangle\lesssim 10^0$, the behavior suggests a relation $F_Q^{3s}\propto \langle \hat{n}\rangle^{\beta_{3s}}$ with an increasing exponent $\beta_{3s}$ the larger $\alpha$ but dependent on the occupation. Beyond this region, the scaling of their QFI is largely suppressed leading to a SQL or worse, depending on $\alpha$. In this regime, the pump field begins to be unable to provide more excitations to the trisqueezed local oscillator, ultimately leading to Rabi oscillations between them. However, thanks to the large improvement at lower occupations, $F_Q^{3s}$ can still be significantly larger than that of standard squeezed states $F_Q^S$ even when $F_Q^{3s}$ scales poorly. For example, for $\langle \hat{n}\rangle\approx 2$, the scaling of $F^{3s}_Q$ is clearly worse than SQL, while providing almost an order of magnitude improvement over $F_Q^S$.  The QFI of  quartsqueezed states $F_Q^{4s}$ reveals a similar behavior to $F_Q ^{3s}$ (cf. Fig.~\ref{fig:QFI_multisqueezed}(b)), albeit shifted to lower occupations. For $\langle \hn \rangle\lesssim 2\cdot 10^{-2}$, it features a SQL, with a larger constant improvement with respect to standard squeezed states, $F^{4s}_Q/F_Q^{S}\approx 2$. Then its scaling exponent $\beta_{4s}$, such that $F_Q^{4s}\propto \langle \hat{n}\rangle^{\beta_{4s}}$, seems again to increase for growing $\alpha$ values in the interval $2\cdot 10^{-2}\lesssim \langle \hat{n}\rangle \lesssim 10^{-1}$. The finite number of excitations provided by the pump field becomes critical at high occupations, where $F_Q^{4s}$ largely depends on $\alpha$, with a reduced scaling, but still allowing for a significant improvement with respect to the optimal Gaussian standard squeezed states $F_Q^S$.

        \begin{figure}
            \centering
            \includegraphics[width=\linewidth]{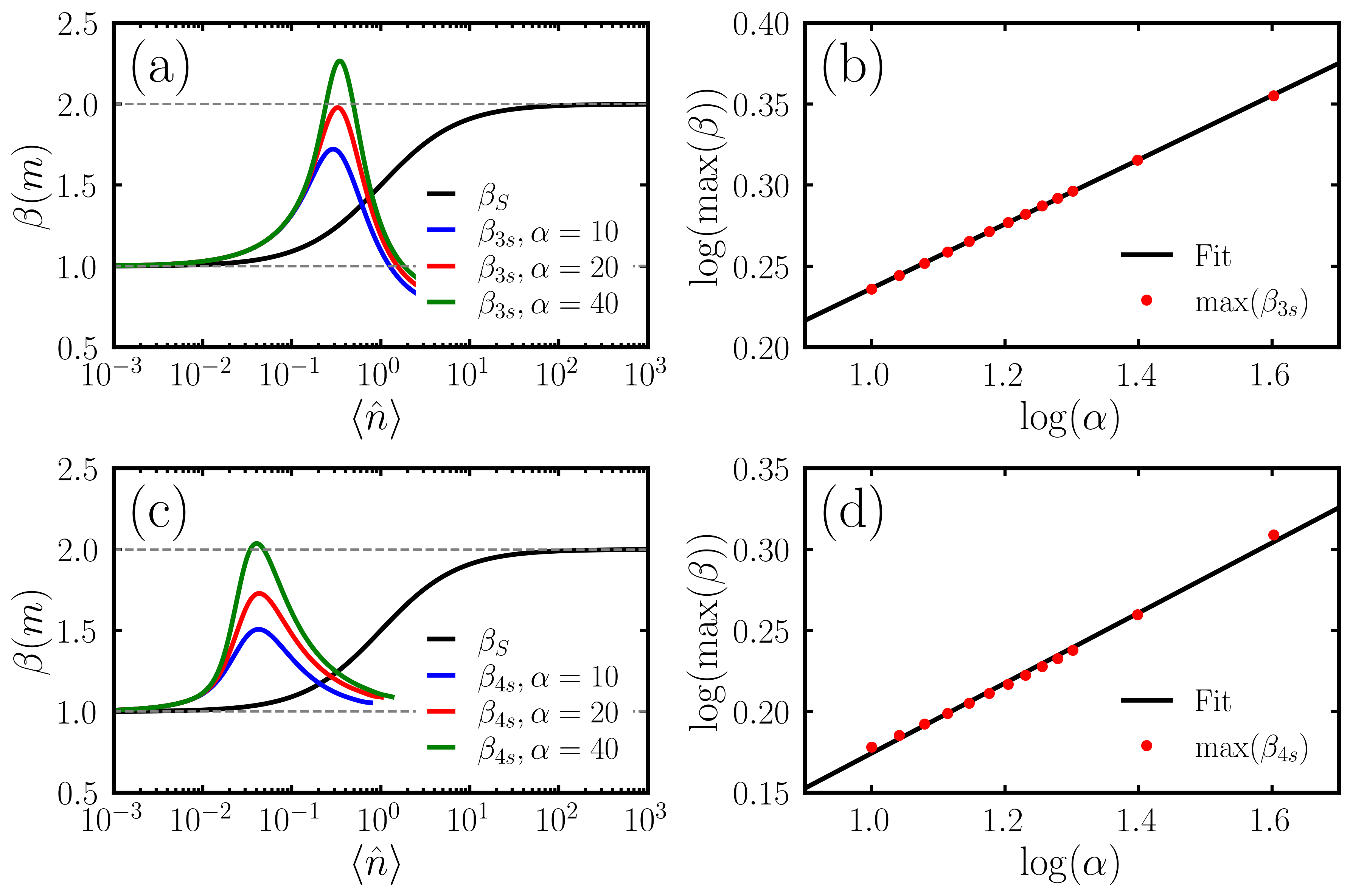}
            \caption{Scaling exponent $\beta_{ms}$ for trisqueezed $m=3$ and quartsqueezed $m=4$ states, in panels (a) and (b), and (c) and (d), respectively.  Panels (a) and (c) show  the numerically-computed scaling exponent $\beta_{ms}$ such that $F_Q^{ms}\propto \langle \hat{n}\rangle^{\beta_{ms}}$ for different initial amplitudes of the pump coherent state $\ket{\alpha}$. For comparison, we include the scaling exponent   for standard squeezed states, shown with a solid black line that ranges from the SQL ($\beta=1$) to the Heisenberg limit ($\beta=2$). Panels (b) and (d) show the maximum scaling exponent $\max\beta_{ms}$ for $m=3$ and $m=4$, respectively, as a function of the initial amplitude $\alpha$. The points correspond to the numerical results $\log_{10}(\max\beta_{ms})$ versus $\log_{10}\alpha$, both in base $10$, while the solid line is the best fit to Eq.~\eqref{eq:fit_max_diffQFI}, leading to the coefficients given in the main text. }
            \label{fig:dlogQFI_multisqueezed}
        \end{figure}

        At intermediate occupations, these multisqueezed states feature a growing scaling exponent $\beta_{ms}$ with increasing $\alpha$, such that $F_Q^{ms}\propto \langle \hat{n}\rangle^{\beta_{ms}}$ and dependent on $\langle \hat{n}\rangle$. In order to quantify this scaling, we numerically compute $\beta_{ms}=\frac{d\log_{10} F_Q^{ms}}{d\log_{10}\langle \hat{n}\rangle}$ for both, trisqueezed $m=3$ and quartsqueezed $m=4$ states. The results for the scaling exponent are shown in Fig.~\ref{fig:dlogQFI_multisqueezed}. In particular, Fig.~\ref{fig:dlogQFI_multisqueezed}(a) shows $\beta_{3s}$ as a function of the occupation. At low occupation, it clearly reveals a SQL region, $\beta_{3s}=1$ for $\langle \hat{n}\rangle \lesssim 10^{-2}$. Yet, for an intermediate occupation, $10^{-1}\lesssim \langle \hat{n}\rangle \lesssim 10^0$, the scaling exponent largely depends on the initial amplitude of the coherent state $\alpha$ of the pump. For an increasing $\alpha$, the exponent $\beta_{3s}$ grows, even surpassing a Heisenberg-limited precision at some occupations, e.g. $\beta_{3s}>2$ for $\alpha=40$ at $\langle \hat{n}\rangle\approx 4\cdot 10^{-1}$.  Moreover, $\beta_{3s}$ seems to increase without an apparent limit. Since numerical computation is restricted to finite values of $\alpha$, we analyze the dependence of the maximum of $\beta_{3s}$ for different fixed values of $\alpha$ and then fit them according to
        \begin{align}\label{eq:fit_max_diffQFI}
            \log_{10} \left(\max \beta_{ms}\right)= a_{ms}\cdot \log_{10}\alpha +b_{ms}.
        \end{align}
        The fitted curve shows an excellent agreement with the computed $\max\beta_{3s}$ values  (cf. Fig.~\ref{fig:dlogQFI_multisqueezed}(b)), where $a_{3s}=0.1984(4)$ and $b_{3s}=0.0379(6)$. This provides a strong numerical evidence that $F_Q^{3s}$ grows indefinitely for increasing $\alpha$ in the region $10^{-1}\lesssim \langle \hat{n}\rangle \lesssim 10^0$, being its scaling exponent $\beta_{3s}$ increasingly larger. We find similar results for quartsqueezed states, but as commented above, their behavior is shifted to lower occupations. In Fig.~\ref{fig:dlogQFI_multisqueezed}(c), the numerically computed scaling exponent $\beta_{4s}$ reveals again a strong dependence on $\alpha$ at $10^{-2}\lesssim \langle \hat{n}\rangle \lesssim 10^{-1}$, that momentarily surpasses $2$ for $\alpha=40$. The numerical fit of $\max\beta_{4s}$ to  Eq.~\eqref{eq:fit_max_diffQFI} yields  the coefficients $a_{4s}=0.216(4)$ and $b_{4s}=-0.042(5)$, showing an excellent agreement with the computed $\max\beta_{4s}$ (cf. Fig.~\ref{fig:dlogQFI_multisqueezed}(d)). These numerical results strongly suggest that multisqueezed states with $m\geq 3$ feature a diverging QFI  when the pump field is able to provide infinitely many excitations to produce the $m$-photon down conversion, i.e. $\lim_{\alpha\rightarrow \infty}F_Q^{ms}\rightarrow \infty$ at some finite occupation, leading also to a divergent scaling exponent $\beta_{ms}\rightarrow \infty$. Note that this is also in agreement with the divergence found in multisqueezed states when defined according to Eq.~\eqref{eq:xim}~\cite{Gordillo2026}, as they stem from a classical approximation of the pump field ($\alpha\rightarrow \infty$)~\cite{Hillery1990}.

        \changes{Although for this family we cannot obtain an analytical scaling relation of the QFI in terms of $m$ at fixed occupation, as we did for \textit{m}th-phase states, we can state that the QFI increases with $m$ for a sufficiently high $\alpha$. This can be deduced since the point where the QFI of these states starts to increase indefinitely with the pump amplitude $\alpha$, occurs for smaller occupations as we increase $m$.}

\section{Useful metrological advantage}\label{s:adv}

    As we have discussed in Sec.~\ref{s:NGstates}, the two distinct families of high-order squeezed states yield larger QFI than standard squeezed states at a fixed occupation. Yet, it is worth  noting that a larger QFI does not always imply a \textit{useful} metrological advantage~\cite{Braunstein1992}. That is, although the previously reported theoretical advantage demonstrates  that measuring an optimal observable leads to an improved precision, any useful advantage must take into account whether such improvement survives when restricting the set of possible observables. This is especially relevant for non-Gaussian states, where a large metrological advantage may come at the expense of measuring increasingly intricate and experimentally demanding observables.

    In order to quantify the metrological usefulness of such families of non-Gaussian states, we introduce the sensitivity $S$ to obtain information with respect to the target parameter $\theta$ when measuring an observable $\hat{A}$, i.e.
    \begin{align}\label{eq:S}
        S(\hat{A}(\theta)) = \frac{1}{(\Delta\hat{A}(\theta))^2} \Biggl( \frac{d\langle \hat{A}(\theta + \delta \theta) \rangle}{d\delta\theta} \Bigg|_{\delta \theta = 0} \Biggr)^2,
    \end{align}
    where $(\Delta\hat{A}(\theta))^2=\langle \hat{A}^2(\theta)\rangle-\langle \hat{A}(\theta)\rangle^2$ denotes the variance of the observable $\hat{A}$ over the state $\rho_\theta$ that carries the information about $\theta$, and $\langle \hat{A}(\theta)\rangle={\rm Tr}[\hat{\rho}_\theta \hat{A}]$ the expectation value. 

    
    Although the QFI of standard squeezed states $F_Q^S$ can be saturated when measuring $\hx^2$ or $\hp^2$ observables, non-Gaussian states require in general complicated higher-than quartic observables. For this reason, we consider a restricted pool of observables $\{\hat{A}_n\}$ and then find the best linear combination $\hat{M}=\sum_n c_n \hat{A}_n$ such its sensitivity is maximized. This can be done as follows~\cite{Fadel_2025}. Given $\{\hat{A}_n\}$, the linear combination of this set $\hat{M} = \sum_n c_n \hat{A}_n$ that maximizes the sensitivity of a measure using the state $\hat{\rho}_\theta$, comes from $c \propto G^{-1} q$ \cite{Fadel_2025}. Here, $q$ is a vector of components $q_n = -i \langle [\hat{A}_n, \hat{H}] \rangle$, with $\hat{H}$ the generator of the rotation, $\hat{H}=\hadaga$; and $G^{-1}$ represents the inverse of the covariance matrix, with matrix elements $G_{n,m} = \frac{1}{2} \langle ( \hat{A}_n \hat{A}_m + \hat{A}_m \hat{A}_n ) \rangle - \langle \hat{A}_n \rangle \langle \hat{A}_m \rangle$, where the expectation value is taken over the state $\hat{\rho}_\theta$.

    In the following we construct the optimal linear combination $\hat{M}$, and analyze its corresponding sensitivity for both families of high-order squeezing, employing a set of observables comprising symmetrized combinations of operator $\hx$ and $\hp$ up to a certain order $k$. We refer also to the App.~\ref{app:adv} for a more restrictive case where a sub-optimal fixed non-Gaussian observable is employed.

    \subsection{\textit{m}th-phase states sensitivity} \label{s:mphase_sensitivity}
        
\begin{figure}
            \centering
            \includegraphics[width=0.8\linewidth]{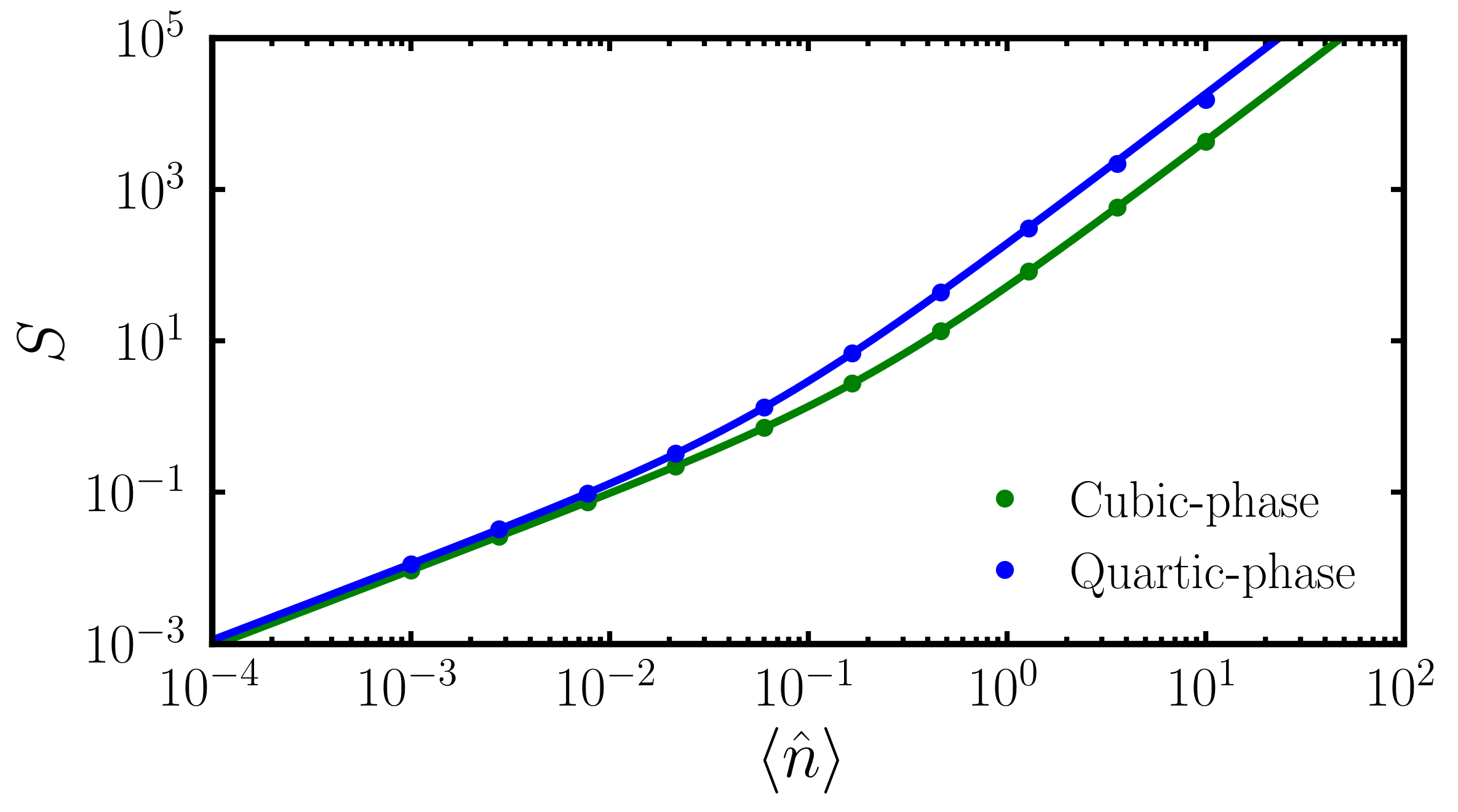}
            \caption{Sensitivity $S(\hat{M})$ (points) as a function of the occupation $\langle \hat{n}\rangle$ for cubic- (green) and quartic-phase (blue) states. For comparison, we include the QFI for these states, $F_Q^{mp}$, shown with solid lines (cf.  Eq.~\eqref{eq:FQm}). The optimal observable $\hat{M}$ is constructed from all symmetrized combinations of the operators $\hx$ and $\hp$ up to fourth and sixth order, for cubic- and quartic-phase states, respectively. The points have been numerically computed using a truncated Fock basis with $N = 10^4$ elements. See main text for details.}
            \label{fig:S_mphase}
        \end{figure}

        As in Sec.~\ref{s:NGstates}, we focus on cubic- and quartic-phase states. First of all, we note that cubic-phase states $\ket{\gamma_3}$ (cf. Eq.~\eqref{eq:gammam}) have been recently analyzed in~\cite{3ps_Gessner_2025} although involving standard squeezing as well. This recent work has shown that a pool of observables containing symmetrized combinations of operators $\hx$ and $\hp$ up to order $k=4$ is enough to construct an optimal observable $\hat{M}$ that saturates their corresponding QFI, $F_Q^{3p}$. In particular, the pool up to fourth order reads as $\{\hat{A}_n\}=\{ \hat{x},\hat{p},\hat{x}^2,\frac{1}{2}(\hat{x}\hat{p}+\hat{p}\hat{x}),\hat{p}^2, \hat{x}^3,\frac{1}{3}(\hat{x}^2\hat{p}+\hat{x}\hat{p}\hat{x}+\hat{p}\hat{x}^2),\frac{1}{3}(\hat{x}\hat{p}^2+\hat{p}\hat{x}\hat{p}+\hat{p}^2\hat{x}),\hat{p}^3,\hat{x}^4,\frac{1}{4}(\hat{x}^3\hat{p}+\hat{x}^2\hat{p}\hat{x}+\hat{x}\hat{p}\hat{x}^2+\hat{p}\hat{x}^3),\frac{1}{6}(\hat{x}^2\hat{p}^2+\hat{x}\hat{p}\hat{x}\hat{p}+\hat{x}\hat{p}^2\hat{x}+\hat{p}\hat{x}^2\hat{p}+\hat{p}\hat{x}\hat{p}\hat{x}+\hat{p}^2\hat{x}^2),\frac{1}{4}(\hat{x}\hat{p}^3+\hat{p}\hat{x}\hat{p}^2+\hat{p}^2\hat{x}\hat{p}+\hat{p}^3\hat{x}),\hat{p}^4\}$.  We numerically calculate the optimal operator $\hat{M}$ using a truncated Fock basis with $N = 10^4$ elements for different photon occupations $\langle \hat{n}\rangle$. Using these optimal operators, we find the associated sensitivity $S(\hat{M})$ under a rotation protocol with a state $\ket{\gamma_3}$. The results for the sensitivity are shown in Fig.~\ref{fig:S_mphase} with green dots, together with $F_Q^{3p}$, as a function of $\langle \hat{n}\rangle$. The specific structure of the optimal observable $\hat{M}$, i.e. the linear combination $\hat{A}_n$, changes depending on the occupation. As it can be seen in Fig.~\ref{fig:S_mphase}, and in agreement with Ref.~\cite{3ps_Gessner_2025}, the operators $\hat{M}$ yield a sensitivity that saturates $F_{Q}^{3p}$, both in the SQL ($F_Q^{3p} \propto \langle \hn \rangle$) and  Heisenberg limit regime ($F_Q^{3p} \propto \langle \hn^2 \rangle$). 

        For quartic-phase states, $\ket{\gamma_4}$, we proceed in a similar fashion. However, a higher squeezing order demands a larger order in the pool of observables to find an optimal observable $\hat{M}$ that saturates the QFI $F_Q^{4p}$. By including all symmetrized combinations of $\hx$ and $\hp$ up to sixth order $k=6$ in $\{\hat{A}_n\}$, the sensitivity when measuring the different observables $\hat{M}$ reaches $F_Q^{4p}$, both in the SQL and Heisenberg limit regimes. The results are shown in Fig.~\ref{fig:S_mphase}, where the blue dots correspond to $S(\hat{M})$, while the solid line depicts $F_Q^{4p}$. The small deviation of the sensitivity at $\langle \hat{n}\rangle=10$ stems from the Fock-basis truncation ($N=10^4$) when computing $\hat{M}$ as it involves  sixth-order operators in $\hx$ and $\hp$.

        Therefore, in order to achieve a useful metrological advantage employing $m$-th phase states and saturate their corresponding QFI well beyond standard squeezed states, one needs to measure observables that contain contributions of $\hx$ and $\hp$ up to order $k=4$ and $k=6$ for cubic-phase and quartic-phase, respectively. \changes{Besides, numerical results suggest that, as we increase $m$, the saturation of the QFI demands a pool with observables up to an order $k=2(m-1)$.}


        



    \subsection{Multisqueezed states sensitivity}\label{s:multi_sensitivity}

\begin{figure}
            \centering
            \includegraphics[width=\linewidth]{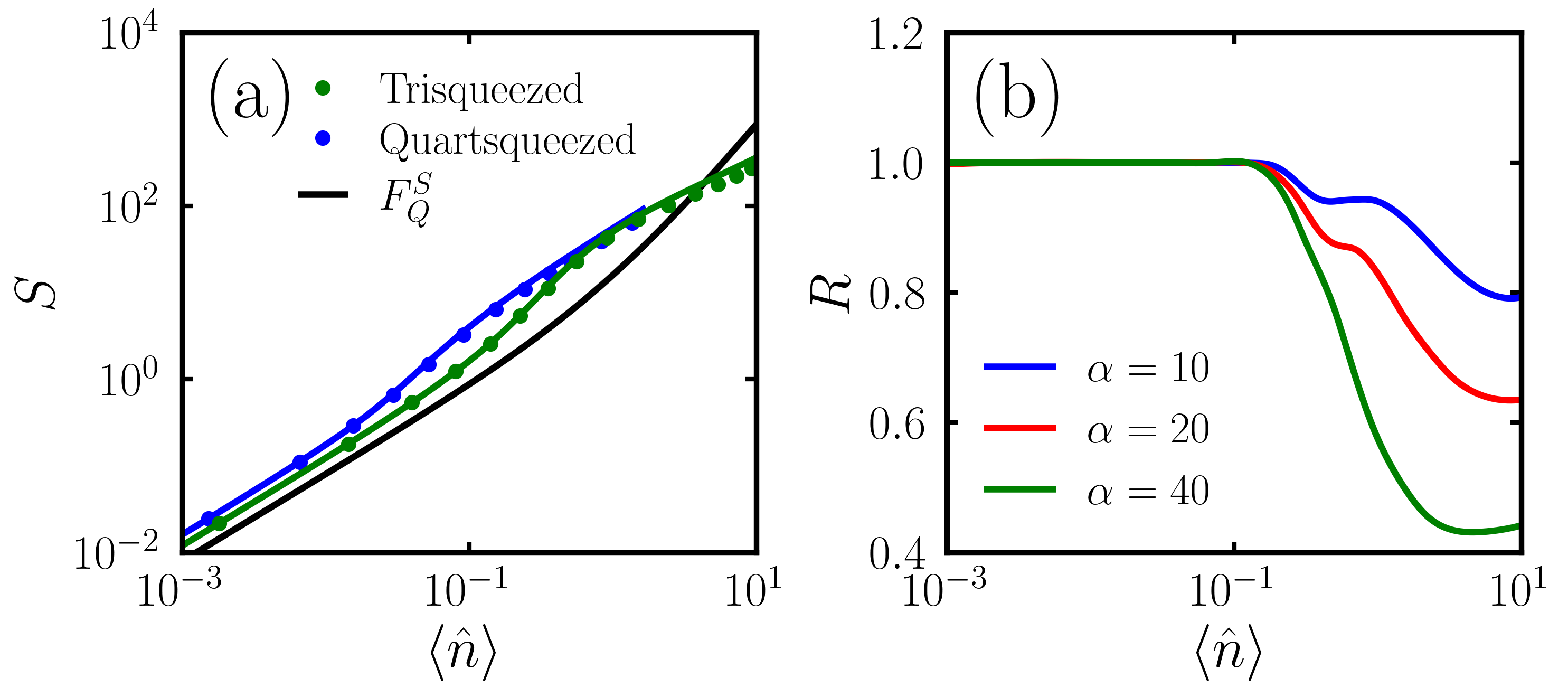}
            \caption{(a) Sensitivity $S(\hat{M})$ (points) as a function of the occupation $\langle \hat{n}\rangle$ for trisqueezed (green) and quartsqueezed (blue) states with $\alpha=10$, together with their corresponding QFI, i.e. $F_Q^{3s}$ and $F_Q^{4s}$ (solid lines). We also include the standard squeezed state QFI, $F_Q^S$, for comparison (black solid line). The optimal observable $\hat{M}$ is built from all symmetrized combinations of the operators $\hx$ and $\hp$ up to sixth \changes{($k=6$)} and twelfth \changes{($k=12$)} order, for trisqueezed and quartsqueezed states, respectively.  (b) Ratio of the sensitivity to QFI of the trisqueezed states, $R=S(\hat{M})/F_Q^{3s}$, for different initial amplitudes of the coherent state $\alpha$, \changes{and with $k=6$}. It reveals a decreasing $R$ as $\alpha$ grows. To ensure numerical convergence, in all cases the population of multisqueezed states at large Fock states is kept below $10^{-17}$. See main text for details. } 
            \label{fig:S_multisqueezed}
        \end{figure}

        In a similar manner, we analyze the sensitivity of trisqueezed and quartsqueezed states by constructing an optimal observable $\hat{M}$ from a pool of observables $\{\hat{A}_n\}$.  In particular, we show here the results for the sensitivity $S(\hat{M})$ when including all symmetrized combinations of $\hx$ and $\hp$ up to sixth \changes{($k=6$)} and twelfth \changes{($k=12$)} order for trisqueezed and quartsqueezed states, respectively. Similar conclusions can be found for larger orders in the pool of observables \changes{(see App.~\ref{app:metrol_adv_ms_k})}, while we refer to App.~\ref{app:adv} for the study of a fixed non-Gaussian measurement. The resulting sensitivities for an initial coherent state $\ket{\alpha=10}$ for the pump field are shown with points in Fig.~\ref{fig:S_multisqueezed}(a). For both cases $m=3$ and $m=4$, we can observe that in the SQL regime, $F_Q^{ms} \propto \langle \hn \rangle$, the sensitivities saturate their QFI, represented as solid lines in the same figure. Indeed, it is not required to consider such high orders of $\hx$ and $\hp$ in $\hat{M}$ to achieve it (see App.~\ref{app:adv}). However, in the occupation regimes where the slope of the QFIs increase indefinitely with $\alpha$, $S(\hat{M})$ is close but does not saturate $F_Q^{ms}$ \changes{(see App.~\ref{app:metrol_adv_ms_k})}. Yet, it is important to stress that $S(\hat{M})$ presents an improvement with respect to the standard squeezed state, thus allowing for a useful metrological advantage. 

        A natural question is whether the sensitivity, albeit not saturating the QFI, also grows with $\alpha$. However, and in contrast to $m$-th phase states, our numerical results show that saturating the QFI for these states beyond the SQL region requires a growing number of observables involving high-order combinations of $\hx$ and $\hp$ as $\alpha$ increases. This is illustrated in Fig.~\ref{fig:S_multisqueezed}(b) for trisqueezed states, where we show the ratio of $R=S(\hat{M})/F_Q^{3s}$ as a function of the occupation for different values of the initial amplitude of the coherent pump field, i.e. $\alpha=10$, $20$ and $40$. For each case, we construct the optimal observable $\hat{M}$ from the pool $\{\hat{A}_n\}$ up to sixth order in $\hx$ and $\hp$. There we can observe that, in the SQL regime, the sensitivities match the QFI for every $\alpha$ value, i.e. $R=1$. However, in the occupation regime where the slope of the QFI increases indefinitely with $\alpha$, $\langle \hat{n}\rangle\gtrsim 10^{-1}$, the ratio $R$  decreases. This decrease stems from a growing QFI, as shown in Sec.~\ref{s:NGstates}, while $S(\hat{M})$  remains approximately constant. To further increase the sensitivity, a larger set of observables would be needed, that is, higher-order symmetrized combinations of $\hx$ and $\hp$. This, however, will increase the complexity and the computational cost of the problem. Although not explicitly shown, we note that the results for quartsqueezed states are equivalent. In this manner, despite trisqueezed and quartsqueezed states feature an increasing QFI at intermediate occupations for a growing $\alpha$, harnessing their full metrological advantage demands the measurement of an increasingly intricate observable involving high-order combination of $\hx$ and $\hp$. Even so, the sensitivities at a fixed occupation can be significantly larger than the optimal Gaussian state (cf. Fig.~\ref{fig:S_multisqueezed}(a)), i.e. the standard squeezed state.

\section{Impact of decoherence channels}\label{s:decoh}

    Finally, we turn our attention to decoherence effects on the potential metrological advantage of these high-order squeezed states. Given that any quantum system is unavoidably affected by noise~\cite{Breuer}, any practical and useful metrological protocol must take into account its robustness against different decoherence mechanisms. Here, we focus on two relevant and standard decoherence channels, namely, pure dephasing and zero-temperature damping noise, and analyze how the QFI is modified by the strength of the considered channel. \changes{Besides, although these channels describe environmental noise acting during the sensing protocol, they also provide qualitative insight into the robustness of the metrological advantage against generic imperfections, such as preparation errors, that would reduce the state coherence and/or introduce mixing. } Both channels are characterized by a complete-positive and trace preserving map such that $\Phi_s: \hat{\rho}\rightarrow \Phi_s(\hat{\rho})$, so that the initial state $\hat{\rho}$ is transformed into another $\Phi_s(\hat{\rho})$ depending on the decoherence strength $s\geq 0$. Note that for $s=0$, the channel is simply the identity, $\Phi_{s=0}(\hat{\rho})=\hat{\rho}$, while for $s>0$, the output state undergoes decoherence. 


    \subsection{Pure dephasing}\label{ss:deph}


        A pure dephasing channel with strength $s$ is characterized by a complete-positive trace-preserving map $\Phi^{\rm pd}_{s}(\hat{\rho})$ that transforms the input density matrix elements $\rho_{n,m}=\bra{n}\hat{\rho}\ket{m}$ according to $\tilde{\rho}_{n,m}=\bra{n}\Phi^{\rm pd}_s(\hat{\rho})\ket{m}$ such that (see App.~\ref{app:decoherence_channels})
        \begin{align}\label{eq:pure_deph}
            \tilde{\rho}_{n,m}=e^{-\frac{1}{2}s (n-m)^2}\rho_{n,m},
        \end{align}
        where $s\geq 0$ (see for example Refs.~\cite{Arqand2020,Mele2024,Leviant2022}). Clearly, pure dephasing channel does not change the energy of the state, namely, ${\rm Tr}[\hadaga \Phi^{\rm pd}_{s}(\hat{\rho})]={\rm Tr}[\hadaga \hat{\rho}]$, but the coherence of $\hat{\rho}$ in the Fock basis is exponentially suppressed with $s$. Therefore, owing to the nature of both families of high-order squeezing states, it is expected that they are more affected by pure dephasing than standard squeezed states.

\begin{figure}
            \centering
            \includegraphics[width=\linewidth]{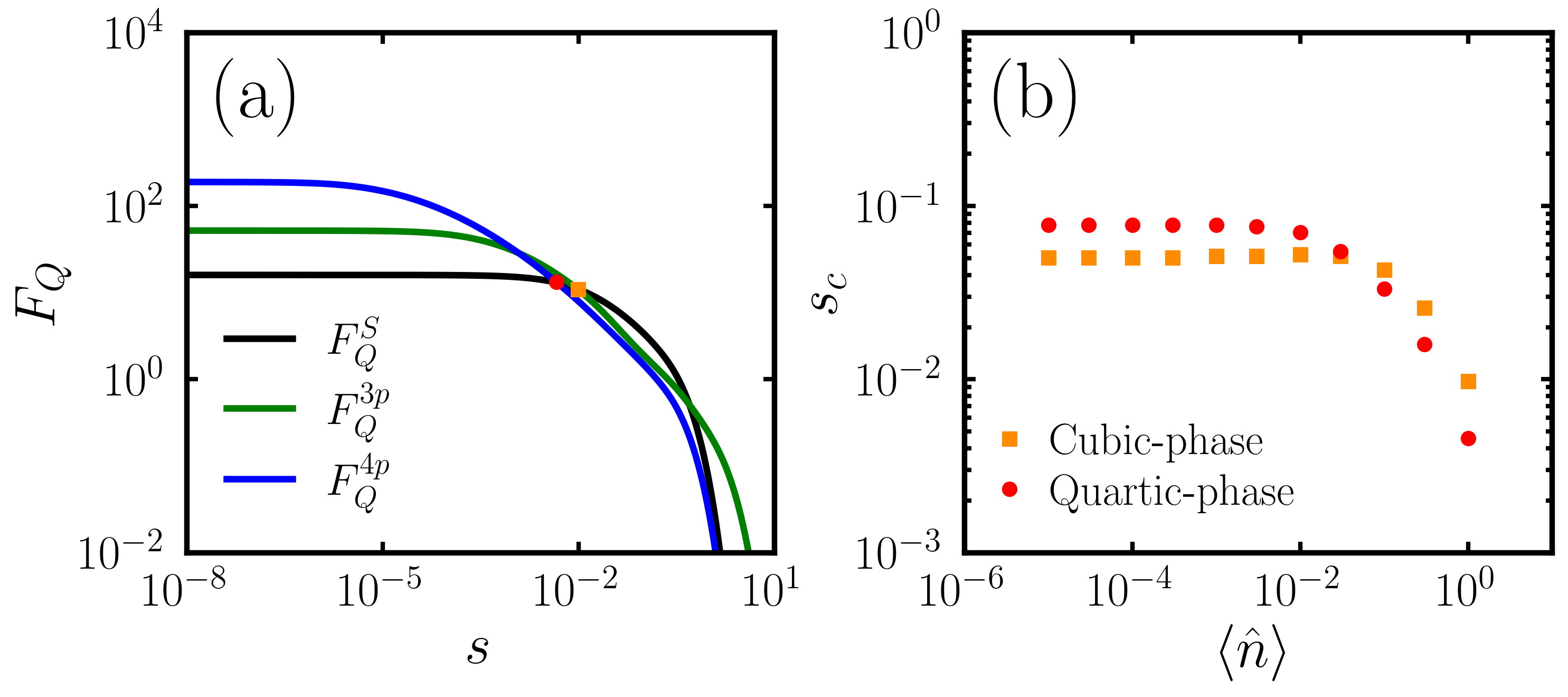}
            \caption{(a) Impact of a pure dephasing channel in the QFI as a function of the strength $s$ for cubic- (green), quartic-phase (blue) and standard squeezed state (black). The results correspond to a fixed occupation $\langle \hn \rangle = 1$. Points mark the position $s_c$ at which $m$-th phase states feature a QFI below dephased standard squeezed states. (b) Value of the critical pure dephasing strength $s_c$ as a function of the occupation $\langle \hat{n}\rangle$ for cubic- (orange squares) and quartic-phase (red circles) states. Numerical results were computed using a truncated Fock basis with $N=10^2$ elements if $\langle \hn \rangle \le 0.1$ and $N=10^3$ if $\langle \hn \rangle > 0.1$.}
            \label{fig:pure-dephasing_mphase}
        \end{figure}

        In order to quantify the robustness of these states, we compute the resulting QFI under the channel $\Phi^{\rm pd}_s(\hat{\rho})$ at a fixed occupation and as a function of $s$, and compare them to the resulting QFI of standard squeezed states under same conditions. As expected, we observe a stronger degradation of their QFI than dephased standard squeezed states. This can be observed in Fig.~\ref{fig:pure-dephasing_mphase}(a), where we show the resulting QFI for cubic- and quartic-phase states as a function of $s$ for a fixed occupation $\langle \hat{n}\rangle=1$. The comparison with standard squeezed states reveals that there exists a finite value $s_c$ at which the QFI of these non-Gaussian states falls below that of dephased standard squeezed states. The points in Fig.~\ref{fig:pure-dephasing_mphase}(a) correspond to $s_c$ for each case. Note that, although for even larger values of $s$,  $F_Q^{3p}$ and $F_Q^{4p}$ are again larger than $F_Q^S$, this is an uninteresting region as the states convey almost no information about $\theta$. In this manner, the potential metrological advantage is lost under pure dephasing when $s\geq s_c$. This behavior persists for any occupation \changes{and, although not explicitly shown, similar conclusions can be drawn when studying the effect of this channel in the sensitivity of the observable $\hat{M}$ computed in the previous section}. 
        
        In Fig.~\ref{fig:pure-dephasing_mphase}(b) we show $s_c$ as a function of $\langle \hat{n}\rangle$, again for cubic- and quartic-phase states. In the SQL regime, the critical value $s_c$ remains approximately constant. We recall that the SQL region takes place at $\langle \hn \rangle \lesssim 10^{-1}$ and $\langle \hn \rangle \lesssim 10^{-2}$ for cubic- and quartic-phase states, respectively. Note that $s_c$ is larger for quartic-phase states. This does not mean that quartic- are more robust to dephasing than cubic-phase states. To the contrary, this larger $s_c$ value for quartic-phase states is simply a consequence of the starting QFI at $s=0$. That is, at $s=0$ one has $F_Q^{4p}>F_Q^{3p}$, so the metrological advantage of $F_Q^{4p}$ with respect to $F_Q^S$ holds up to a stronger dephasing channel.  Yet, in the Heisenberg limit regime, as the states acquire larger excitations and larger QFI, they also  become increasingly fragile against dephasing. Although not explicitly shown in the main text, trisqueezed and quartsqueezed behave in analogous manner, leading to equivalents plots as those in Fig.~\ref{fig:pure-dephasing_mphase}. We refer to App.~\ref{app:decoh_ms} for further details on the impact of pure dephasing on multisqueezed states.

    \subsection{Zero-temperature damping}\label{ss:damp}

        Another relevant decoherence mechanism is the so-called zero-temperature or amplitude damping channel. This decoherence channel is characterized by a complete-positive and trace-preserving map $\Phi^{\rm zd}_{s}(\hat{\rho})$~\cite{Chuang1997,Liu04} that transforms the density matrix elements $\rho_{n,m}$ according to (see App.~\ref{app:decoherence_channels})
        \begin{align}\label{eq:zero_damp}
            &\tilde{\rho}_{n,m}=e^{-\frac{s(n+m)}{2}}\nonumber \\ &\times\sum_{k=0}^\infty (1-e^{-s})^k \sqrt{\frac{(n+k)! (m+k)!}{(k!)^2 n!m!}}\rho_{n+k,m+k},
        \end{align}
        with $s\geq 0$ and $\tilde{\rho}_{n,m}=\bra{n}\Phi^{\rm zd}_s(\hat{\rho})\ket{m}$. Again, for $s=0$ the state remains unchanged, while for $s\to \infty$ the state is brought to the vacuum state, $\lim_{s\to\infty}\Phi^{\rm zd}(\hat{\rho})=\ket{0}\bra{0}$. 

\begin{figure}
            \centering
            \includegraphics[width=\linewidth]{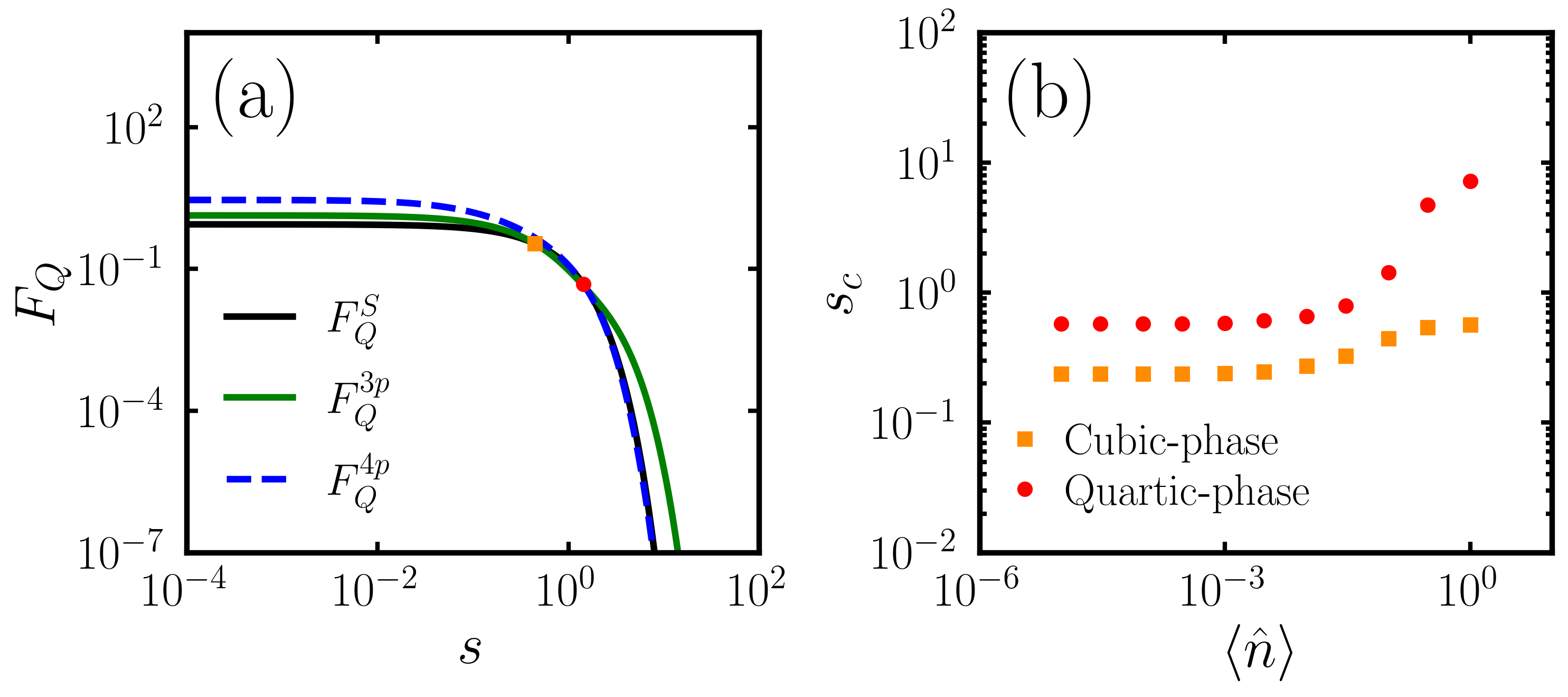}
            \caption{(a) Impact of a zero-temperature damping channel in the QFI of a state as a function of the channel strength $s$ and for a fixed occupation $\langle \hn \rangle = 0.1$. Green line is for cubic-phase states, dashed blue line is for quartic-phase states and black line corresponds to standard squeezed states. (b) Potential metrological advantage limit $s_c$ of the cubic-phase (orange squares) and quartic-phase (red circles) states over the standard squeezed states as a function of the initial photon occupation $\langle \hn \rangle$. Results of both graphs were calculated using a truncated Fock basis with $N=10^2$ elements if $\langle \hn \rangle \le 0.1$ and $N=10^3$ if $\langle \hn \rangle > 0.1$.}
            \label{fig:zeroTdamping_mphase}
        \end{figure}

        We proceed in a similar manner as for pure dephasing, analyzing the robustness of the QFI of high-order squeezing states as a function of the damping strength $s$. The results for a fixed initial occupation $\langle \hat{n}\rangle=0.1$ are shown in Fig.~\ref{fig:zeroTdamping_mphase}(a), which again reveal a critical value $s_c$ above which cubic- and quartic-phase states loose their metrological advantage over standard squeezed states. \changes{Although not explicitly shown, similar results are obtained when studying the effect of this channel in the sensitivity of the observable $\hat{M}$ computed in Sec.~\ref{s:mphase_sensitivity}.} As in pure dephasing noise, for large values of $s$, these states recover the advantage, but it takes places in a poor metrological regime where the states carry almost no information about $\theta$. Yet, in contrast to dephasing, the behavior of the critical value $s_c$ as a function of the occupation indicates that the robustness to zero-temperature damping of these high-order squeezed states compared to that of the standard squeezed states increases with the occupation. As it can be seen in Fig.~\ref{fig:zeroTdamping_mphase}(b), while $s_c$ remains approximately constant in the SQL regime, in the Heisenberg limit region, $s_c$ increases for both, cubic- and quartic-phase states. At larger occupations, the value $s_c$ seems to saturate, which can be attributed to a constant robustness once standard and high-order squeezed states settle in the Heisenberg limit, $\langle \hat{n}\rangle\gtrsim 1$. In addition, we note that quartic-phase states display a larger critical value $s_c$ thanks to their greater metrological advantage at $s=0$. Thus, their QFI remains above that of standard squeezed states up a stronger zero-temperature noise channel. The analysis for trisqueezed and quartsqueezed states lead to equivalent results and conclusions, and therefore we refer to the interested reader to App.~\ref{app:decoh_ms} for further details.

\section{Conclusions}\label{s:conc}

    In this article we investigate the metrological performance of two non-Gaussian states families under the standard interferometric or rotation protocol, comparing the results with those for the optimal Gaussian state, i.e. standard squeezed states, under equal occupations. In particular, we consider two nonequivalent families of high-order squeezed states, referred to as \textit{m}th-phase and multisqueezed states. Among these non-Gaussian states, we focus on the lowest non-trivial orders, i.e. $m=3$ and $m=4$, namely cubic- and quartic-phase, as well as trisqueezed and quartsqueezed states.

    First, we focus on the potential metrological advantage, and as we show, both families can improve the QFI of the standard squeezed states. In particular, \textit{m}th-phase states present a greater QFI at any occupation, leading to a Heisenberg limited precision with a prefactor that grows exponentially with the squeezing order $m$. Meanwhile, multisqueezed states of order $m>2$ have to be numerically studied with special care to ensure convergence, requiring to explicitly include the pump field.  Under these conditions, we find that the QFI also surpasses that of the squeezed vacuum for reasonably low occupations, having a region where the QFI grows indefinitely with the initial amplitude of the coherent pump field $\alpha$, displaying a remarkably different behavior with respect to \textit{m}th-phase states.

    To complement these results, we study the useful metrological advantage of these states. From a restricted set of observables, we construct the best linear combination such the sensitivity of the measurement is maximized. In this manner, we observe that operators made with the symmetrized combinations of observables $\hx$ and $\hp$ up to orders $k = 4$ and $k=6$ are able to saturate the sensitivity of cubic- and quartic-phase, respectively, for any occupation. Regarding multisqueezed states, we consider operators made with symmetrized combinations of observables $\hx$ and $\hp$ up to orders $k=6$ and $k=12$ for trisqueezed and multisqueezed states, respectively. We find that, while in the SQL regime the sensitivities saturate to their QFI, in the region where the slope of the QFI increases indefinitely with $\alpha$, the sensitivities do not reach their QFI. Yet, the resulting sensitivity improves the QFI  of the standard squeezed states in both cases.  

    Finally, we investigate the robustness of these two families of states to standard decoherence channels, i.e. pure dephasing and zero-temperature damping. Employing standard squeezing as reference state, we find that their metrological advantage is fragile against dephasing, which worsens with increasing photon occupation. To the contrary, these states show a reasonable robustness against damping.  

    Our results show the metrological performance of two distinct families of non-Gaussian states, that stem from high-order squeezing generalizations of coherent and squeezed states. Our work highlights the suitability and limitations of these states to yield a metrological advantage, opening the door for further studies on enhanced metrological protocols harnessing non-Gaussian states.

\begin{acknowledgments}
We acknowledge financial support form the Spanish Government via the project PID2024-161371NB-C21 (MCIU/AEI/FEDER, EU) and project TSI-069100-2023-8 (Perte Chip-NextGenerationEU), as well as the Ram{\'o}n y Cajal (RYC2023-044095-I) and (RYC2020-030060-I) research fellowships. 
\end{acknowledgments}

\appendix
\section{Quantum Fisher information of squeezed and coherent states}\label{app:a}
In the following we recall the QFI expressions for squeezed and coherent states under a rotation protocol. Coherent states are defined as $\ket{\alpha}=\hat{D}(\alpha)\ket{0}$, where $\hat{D}(\alpha)=e^{\alpha \hadag-\alpha^*\ha}$. Using the well-known relation    $\hat{D}^\dagger (\alpha) \ha \hat{D}(\alpha)=\hat{a}+\alpha$,
it follows that  $\langle \hn \rangle=\bra{\alpha}\hadaga\ket{\alpha}=|\alpha|^2$. 
Next, we find
\begin{align}
    \langle \hn^2\rangle=\bra{0}\hat{D}^\dagger(\alpha) (\hadaga)(\hadaga)\hat{D}(\alpha)\ket{0}=|\alpha|^4+|\alpha|^2.
\end{align}
Therefore, the QFI of coherent states in terms of the average occupation reads as
\begin{align}
    F_Q^C=4\left(\langle \hn^2\rangle-\langle \hn \rangle^2 \right)=4\langle \hn \rangle.
\end{align}
In a similar manner, for squeezed states, $\ket{\xi}=\hat{S}(\xi)\ket{0}$ with $\hat{S}(r)=e^{1/2(\xi^*\ha^2-\xi \ha^{\dagger,2})}$, we have $\hat{S}^\dagger (\xi)\ha \hat{S}(\xi)=\ha \cosh|\xi|-e^{i\theta}\hadag \sinh|\xi|$. This leads to $\bra{0}\hat{S}^{\dagger}(\xi)\ha^\dagger \ha \hat{S}(\xi)\ket{0}=\sinh^2|\xi|$ and $\bra{0}\hat{S}^{\dagger}(\xi)(\hadaga)^2\hat{S}(\xi)\ket{0}=\sinh^4|\xi|+2\cosh^2|\xi|\sinh^2|\xi|$. In this manner, the QFI under this rotation protocol for squeezed states reads as
\begin{align}\label{eq:fqsque}
    F_Q^S=8(\langle \hn \rangle^2+\langle \hn \rangle).
\end{align}
Finally, let us remark that a general displaced and squeezed state, $\ket{\alpha,\xi}=\hat{D}(\alpha)\hat{S}(\xi)\ket{0}$,  does not improve the QFI above Eq.~\eqref{eq:fqsque} for a given occupation number $\langle \hn \rangle$, that is, the squeezed vacuum is the optimal Gaussian state for a fixed $\langle \hn \rangle$~\cite{Matsubara2019}.

\section{Quantum Fisher information of \textit{m}th-phase states}\label{app:mth-phase}
    In the following we provide the details to obtain the QFI for the \textit{m}th-phase states, given in Eq.~\eqref{eq:FQm} of the main text. These states are defined as 
    \begin{align}
        \ket{\gamma_m}=\exp{\biggl\{\frac{i\gamma_m (\ha+\hadag)^m}{2^{m/2}}}\biggr\}\ket{0} = e^{i\gamma_m \hx^m}\ket{0},
    \end{align}
    where $m \in \mathbb{N}$, $\gamma_m\in\mathbb{R}$  acts as the \textit{m}th-order squeezed parameter, and $\hx$ is the dimensionless position operator, $\hx = \frac{1}{\sqrt{2}}(\hadag + \ha)$. Therefore, we can easily obtain the QFI by switching to the position representation and calculating the average values of $\hn = \hadag \ha$ and $\hn^2 = (\hadaga)^2$. In the following, we consider $m\geq 3$. 

    The number operator, $\hn = \hadaga$, can be expressed in position-momentum representation as
    \begin{align}
        \hn = \frac{1}{2}(\hx^2+\hp^2-1)=\frac{1}{2}(\hx^2-\partial^2_{\hat{x}}-1),
    \end{align}
    since $\ha = \frac{1}{\sqrt{2}}(\hx+i\hp)$ and $\hadag= \frac{1}{\sqrt{2}}(\hx - i\hp)$, and $\hp = -i\frac{\partial}{\partial \hx} \equiv -i\partial_{\hx}$.  The average occupation number reads then as
    \begin{align}\label{eq:n_integral}
        \bra{\gamma_m} \hn \ket{\gamma_m} =   &\frac{1}{2} \int^{+\infty}_{-\infty} dx \langle 0|x \rangle \bra{x}e^{-i\gamma_m \hx^m} \nonumber \\&\times (\hx^2 - \partial_{\hx}^2 -1) e^{i\gamma_m\hx^m}\ket{x} \langle x|0 \rangle,
    \end{align}
    where $\langle 0|x \rangle$ and $\langle x|0 \rangle$ have the form
    \begin{align}
        \langle 0|x \rangle = \langle x|0 \rangle = \frac{1}{\pi^{1/4}} e^{-x^2/2}.
    \end{align}
    The integral in Eq.~\eqref{eq:n_integral} can be solved employing
    \begin{align}\label{eq:aux_integral}
        \int^{+\infty}_{-\infty} &dx e^{-x^2} x^m =\nonumber \\& =
        \begin{cases}
        0, & \text{if $m$ is odd}, \\\displaystyle \sqrt{\pi} \frac{(m-1)!!}{2^{m/2}}, & \text{if $m$ is even},
        \end{cases}
    \end{align}
    which finally leads to
    \begin{align} \label{eq:avg_n_app}
        \langle \hn \rangle = m^2 \gamma_m^2 \frac{(2m-3)!!}{2^m},
    \end{align}
    where $(m)!!$ denotes the double factorial of $m$. 

   The average squared occupation number $\hn^2$ can be obtained in a similar manner. First, $\hn^2$ in  position representation takes the following form
    \begin{align}
        \hn^2 = \frac{1}{4}(\hx^4 + \partial_{\hx}^4 + \hx^2\partial_{\hx}^2 + \partial_{\hx}^2\hx^2 - 2\hx^2 - 2\partial_{\hx}^2 + 1).
    \end{align}
    The expectation value follows then from
    \begin{align}
        \bra{\gamma_m} \hn^2 \ket{\gamma_m} = & \frac{1}{2} \int^{+\infty}_{-\infty} dx \langle 0|x \rangle \bra{x}e^{-i\gamma_m\hx^m}\nonumber \\&\times \hn^2 e^{i\gamma_m\hx^m}\ket{x} \langle x|0 \rangle.
    \end{align}
    Using again Eq.~\eqref{eq:aux_integral}, we find
    \begin{align}\label{eq:avg_n2_app}
        \langle \hn^2\rangle = &m^4 \gamma_m^4 \frac{(4m-5)!!}{2^{2m}} \nonumber \\&+ m^2 \gamma_m^2 \frac{(2m-5)!!}{2^m} (m^2-2).
    \end{align}

    Since we are interested in the QFI in terms of the average occupation, we introduce Eq.~\eqref{eq:avg_n_app} into~\eqref{eq:avg_n2_app}, to obtain $\langle \hn^2\rangle$ in terms of $\langle \hn \rangle$, i.e.
    \begin{align}
        \langle \hn^2\rangle = \frac{(4m-5)!!}{((2m-3)!!)^2} \langle \hn \rangle^2 + \frac{m^2-2}{2m-3} \langle \hn \rangle.
    \end{align}
    Finally, introducing this expression into Eq.~\eqref{eq:QFI_general}, we obtain the general expression for the QFI for \textit{m}th-phase states with $m\ge3$ in terms of the occupation $\langle \hn \rangle$:
    \begin{align}
        F_Q^{mp} = 4 \left[ \Biggl(\frac{(4m-5)!!}{(2m-3)!!)^2}-1\Bigg) \langle \hn \rangle^2 + \frac{m^2-2}{2m-3} \langle \hn \rangle \right].\nonumber
    \end{align}
    The previous expression corresponds to Eq.~\eqref{eq:FQm} given in the main text.   


\section{Useful metrological advantage with fixed observables}\label{app:adv}

    In the main text, we show the metrological usefulness of these non-Gaussian states by computing the sensitivity $S$ (see Eq.~\eqref{eq:S}) of measuring an operator $\hat{M}$. This operator is built as the best linear combination over a restricted pool of observables. Here instead we focus on a more conservative scenario, considering the sensitivity offered by measuring just a fixed, yet non-Gaussian, observable. As we show below, even in this case one may achieve a metrological advantage with respect to standard squeezed states. Nevertheless, it is worth  stressing that choosing different observables largely modify the sensitivity. This Appendix simply illustrates  that it is still possible to reach an advantage without the need for building optimal operators, as shown in the main text.     

\begin{figure}
            \centering
            \includegraphics[width=\linewidth]{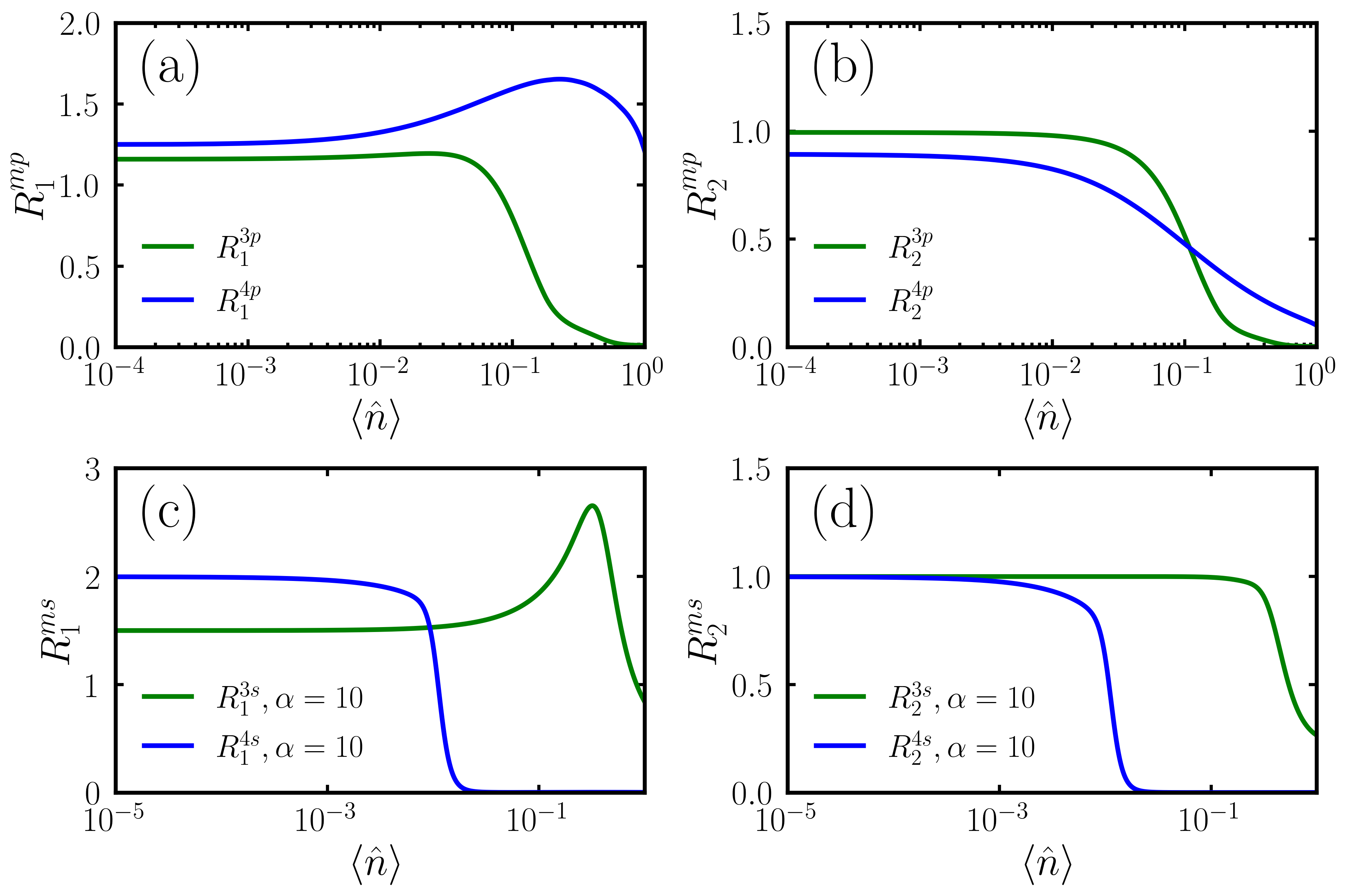}
            \caption{(a) Ratio $R_1^{mp} = S(\hat{B}_{mp})/F^{S}_Q$ between the sensitivity of \textit{m}th-phase states when measuring  $\hat{B}_{mp}$, and the QFI of the standard squeezed states. Green and blue lines correspond to cubic- and quartic-phase states with observables $\hat{B}_{3p} = \hx \hp (\hx + \hp) + \text{h.c.}$ and $\hat{B}_{4p} = \hx^4$, respectively. (b) Ratio $R_2^{mp} = S(\hat{B}_{mp})/F^{mp}_Q$ between \textit{m}th-phase sensitivity  and their corresponding QFI. Same format as in (a).  (c) Ratio $R_1^{ms} = S(\hat{B}_{ms})/F^{S}_Q$ between the multisqueezed states sensitivity measuring the observable $\hat{B}_{ms}$, and the QFI of the standard squeezed states. Green and blue lines correspond to trisqueezed  and quartsqueezed states with $\alpha=10$, respectively, and $\hat{B}_{ms} = \ha^{\dagger, m}+\ha^m$. (d) Ratio $R_2^{ms} = S(\hat{B}_{ms})/F^{ms}_Q$ between the multisqueezed states sensitivity and their QFI, with the same format as in (c).} 
            \label{fig:S_single_op}
        \end{figure}

    Starting with the \textit{m}th-phase states, as in the main text, we focus on cubic- and quartic-phase states. For the quartic-phase state we consider the non-Gaussian observable $\hat{B}_{4p} =\hx^4$ motivated by the structure of these states (cf. Eq.~\eqref{eq:gammam}). However, for the cubic-phase state, the choice $\hx^3$ results in a poor sensitivity, and we employ instead $B_{3p}=\hx \hp (\hx + \hp) + \text{h.c.}$, i.e. a fixed cubic-order combination of $\hx$ and $\hp$. Then we calculate the maximum sensitivity of both operators under a rotation protocol with their respective states, in terms of $\langle \hn \rangle$, and considering a truncated Hilbert space with $N = 10^4$ Fock states. Figure~\ref{fig:S_single_op}(a) shows the ratio $R_1^{mp} = S(\hat{B}_{mp})/F^{S}_Q$ between these obtained sensitivities, and the QFI of the standard squeezed states. In the SQL regime ($\langle \hn \rangle \lesssim 10^{-1}$ and $\langle \hn \rangle \lesssim 10^{-2}$ for cubic- and quartic-phase states, respectively), the maximum sensitivity of both cases is greater than the QFI of the standard squeezed states. Therefore, in this region, the \textit{m}th-phase states are capable of providing an useful metrological advantage over the standard squeezed states measuring these fixed operators. In the Heisenberg limit regime, however, the cubic-phase state looses the advantage, as the ratio $R^{3p}_1$ falls below $1$. Meanwhile, the quartic-phase state ratio $R^{4p}_1$ increases, until it reaches a maximum around $\langle \hn \rangle \approx 0.2$. For further information about these sensitivities, Fig.~\ref{fig:S_single_op}(b) shows the ratio $R_2^{mp} = S(\hat{B}_{mp})/F^{mp}_Q$ between  sensitivity, and their corresponding QFI. For these specific example, the cubic-phase state saturates  its QFI in the SQL, while the quartic-phase state fails to do so at any occupation. 

    For the trisqueezed and quartsqueezed states, we proceed in a similar manner, but with different observables due to their structure. In particular, we consider the non-Gaussian observables $\hat{B}_{ms} = \ha^{\dagger,m} + \ha^m$, with $m = 3$ and $4$, for trisqueezed and quartsqueezed states, respectively. Now, we compute the maximum sensitivity of both operators under a rotation protocol, in terms of $\langle \hn \rangle$, for a fixed $\alpha=10$, and considering a truncated Fock basis such that the population over large Fock states is always below or equal to $10^{-17}$. Similar results can be obtained for other choices of $\alpha$.  Figure~\ref{fig:S_single_op}(c) shows the ratio $R_1^{ms} = S(\hat{B}_{ms})/F^{S}_Q$ between the multisqueezed sensitivity, and the QFI of the standard squeezed states. In the SQL limit regime ($\langle \hn \rangle \lesssim 10^{-1}$ and $\langle \hn \rangle \lesssim 10^{-2}$ for trisqueezed and quartsqueezed states, respectively), the maximum sensitivity of both cases is greater than the QFI of the standard squeezed states. Therefore, in this region, these non-Gaussian states can provide a useful metrological advantage over the standard squeezed states measuring fixed operators. However, in the occupation regime where the slope of the QFI of these states grows indefinitely with $\alpha$, the quartsqueezed state ratio $R^{4s}_1$ abruptly decreases, quickly losing its advantage. For the considered operators, we observe that  the trisqueezed state ratio $R^{3s}_1$ increases significantly, until it reaches a maximum around $\langle \hn \rangle \approx 0.3$, where it starts to fall, ultimately loosing the advantage. As with the \textit{m}th-phase states, for further information about these sensitivities, Fig.~\ref{fig:S_single_op}(d) shows the ratio $R_2^{ms} = S(\hat{B}_{ms})/F^{ms}_Q$ between the sensitivity and their corresponding QFI. For the considered observables, the resulting sensitivity saturates the QFI in the SQL.
    



\section{\changes{Multisqueezed sensitivity as a function of $k$}\label{app:metrol_adv_ms_k}}

\begin{figure}
            \centering
            \includegraphics[width=0.8\linewidth]{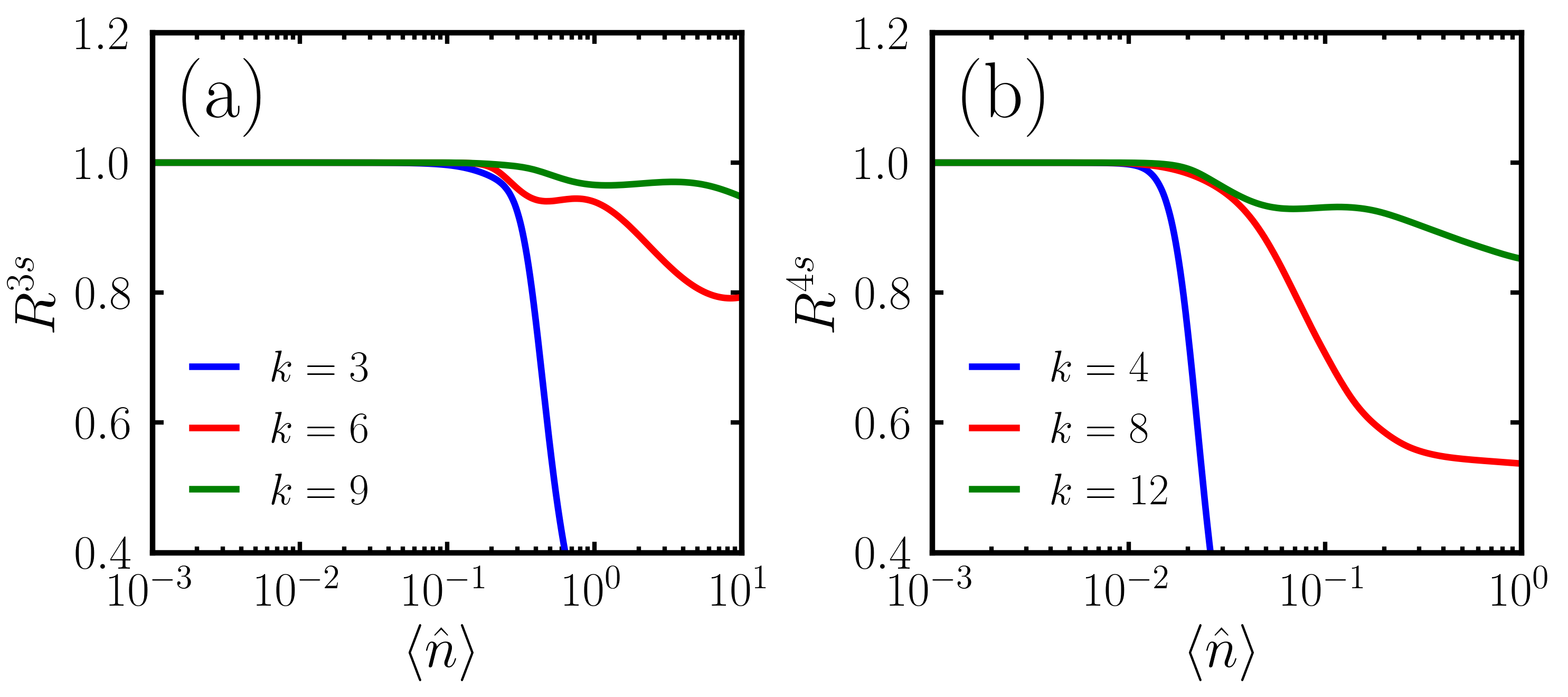}
            \caption{\changes{(a) Ratio of the sensitivity to QFI for trisqueezed states, $R^{3s}=S(\hat{M})/F_Q^{3s}$, as a function of the occupation $\langle \hn \rangle$, for a fixed $\alpha=10$, and considering different maximum orders $k$ in observable $\hat{M}$. (b) Ratio of the sensitivity to QFI for quartsqueezed states, $R^{4s}=S(\hat{M})/F_Q^{4s}$, as a function of the occupation $\langle \hn \rangle$, for a fixed $\alpha=10$, and considering different maximum orders $k$ in observable $\hat{M}$. Both results reveal an increasing $R^{ms}$ as $k$ grows. To ensure numerical convergence, in all cases the population of multisqueezed states at large Fock states is kept below $10^{-17}$.}}
            \label{fig:S_interms_k}
    \end{figure}

    \changes{In the main text, we studied the sensitivity obtained using the observable $\hat{M}$, constructed as the optimal linear combination of all symmetrized combinations of $\hx$ and $\hp$ up to order $k$. Specifically, for trisqueezed and quartsqueezed states, we considered only the cases $k=6$ and $k=12$, respectively. Therefore, in this section we investigate how the sensitivity of these states changes as $k$ is varied, for a fixed $\alpha$. To this end, we numerically compute the ratio $R^{ms}$ between the sensitivity $S(\hat{M})$ and the QFI of the state $F^{ms}_Q$, considering different maximum orders $k \geq m$ in $\hat{M}$, and for $\alpha = 10$. Results are shown in Fig.~\ref{fig:S_interms_k}.}

    \changes{For trisqueezed states, outcomes can be found in Fig.~\ref{fig:S_interms_k} (a). We observe that, in the SQL regime ($F_Q^{ms} \propto \langle \hn \rangle$), the sensitivity saturates the QFI ($R^{3s} = 1)$) for all $k$ considered, as anticipated in the main text. Meanwhile, if we move to the occupation region where the QFI grows indefinitely with $\alpha$ (see main text for details), the sensitivity does not saturate the QFI for any $k$, but it increases as we increase $k$, as expected. Similar conclusions can be drawn from Fig.~\ref{fig:S_interms_k}(b) for quartsqueezed states.}


\section{Decoherence channels} \label{app:decoherence_channels}
    As discussed in the main text, we consider the impact of two standard decoherence processes, namely, pure dephasing and zero-temperature damping. The impact of these decoherence mechanisms can be described in terms of a complete-positive and trace-preserving quantum channel $\Phi_s: \hat{\rho}\to \Phi_s(\hat{\rho})$, that transforms the input state $\hat{\rho}$ into $\Phi_s(\hat{\rho})$, where $s\geq 0$ denotes the noise strength of the channel~\cite{Breuer}. Here we provide a brief description on how pure dephasing and zero-temperature damping can be written as a quantum channel $\Phi_s(\hat{\rho})$, leading to the expressions given in the main text (cf. Eqs.~\eqref{eq:pure_deph} and~\eqref{eq:zero_damp}). 

    Markovian pure dephasing noise acting on a single bosonic mode can be described according to the Lindblad equation with jump operator $\hat{A}=\hadaga$ and rate $\gamma$~\cite{Breuer}, so that
    \begin{align}\label{eq:Lindblad}
        \frac{d}{dt}\hat{\rho}=\gamma\left( \hat{A}\hat{\rho} \hat{A}^\dagger -\frac{1}{2}\left\{\hat{A}^\dagger \hat{A},\hat{\rho} \right\}\right),
    \end{align}
    Upon integration it transforms the initial state $\hat{\rho}$ into $\hat{\tilde{\rho}}$, which can be written in terms of the matrix elements $\rho_{n,m}=\bra{n}\hat{\rho}\ket{m}$ and $\tilde{\rho}_{n,m}=\bra{n}\hat{\tilde{\rho}}\ket{m}$ as
    \begin{align}
        \tilde{\rho}_{n,m}=e^{-\frac{1}{2}\gamma t (n-m)^2}\rho_{n,m}.
    \end{align}
    Thus, defining $s=\gamma t$ as the strength of pure dephasing, one simply finds Eq.~\eqref{eq:pure_deph}~\cite{Arqand2020,Mele2024,Leviant2022}. The new dephased state can be therefore expressed as,
    \begin{align}
        \Phi^{\rm pd}_{s}(\hat{\rho})= \hat{\tilde{\rho}} = \sum_{n,m=0}^{\infty}e^{-\frac{1}{2}s(n-m)^2}\rho_{n,m}\ket{n}\bra{m}.
    \end{align}
    For $s=0$, the quantum channel reduces to the identity, $\Phi_{s=0}^{\rm pd}(\hat{\rho})=\hat{\rho}$, while for $s>0$, the coherence in the $\hadaga$ basis are exponentially suppressed.

    In a similar manner, Markovian zero-temperature damping noise corresponds to Eq.~\eqref{eq:Lindblad} with $\hat{A}=\ha$, and again defining $s$ as the noise strength. The quantum channel for this process can be written as (see for example Ref.~\cite{Liu04})
    \begin{align}
    \Phi^{\rm zd}_s(\hat{\rho})=\sum_{k=0}^\infty \hat{E}_k \hat{\rho} \hat{E}_k^\dagger,    
    \end{align}
    being $\{\hat{E}_k\}$ the Kraus operators, that take the form
    \begin{align}
        \hat{E}_k=\sum_{n=k}^\infty \sqrt{\frac{n!}{k! (n-k)!} \frac{x_s^{n-k}}{e^{s k}}}\ket{n-k}\bra{n},
    \end{align}
    such that $\sum_{k=0}^\infty \hat{E}_k^\dagger \hat{E}_k=\hat{\mathbb{I}}$, and with $x_s=1-e^{-s}$. Therefore, the zero-temperature damping quantum channel reads as
    \begin{align}
        \Phi^{\rm zd}_s(\hat{\rho})=&\sum_{n,m=0}^{\infty} e^{-\frac{s(n+m)}{2}}\sum_{k=0}^\infty x_s^k \nonumber\\ &\times \sqrt{\frac{(n+k)! (m+k)!}{(k!)^2 n!m!}}\rho_{n+k,m+k} \ket{n}\bra{m}.
    \end{align}
    In particular, $\Phi_s^{\rm zd}(\hat{\rho})$ transforms the initial matrix elements $\rho_{n,m}$ into $\tilde{\rho}_{n,m}$ according to 
    \begin{align}
        &\tilde{\rho}_{n,m}=\bra{n}\Phi_{s}^{\rm zd}(\hat{\rho})\ket{m}\nonumber\\&=e^{-\frac{s(n+m)}{2}}\sum_{k=0}^\infty x_s^k \sqrt{\frac{(n+k)! (m+k)!}{(k!)^2 n!m!}}\rho_{n+k,m+k}.
    \end{align}
   The previous expression corresponds to Eq.~\eqref{eq:zero_damp} in the main text. Again, for zero noise strength, $s=0$, $\Phi_{s=0}^{\rm zd}(\hat{\rho})=\hat{\rho}$, while it compresses the state towards $\ket{0}\bra{0}$ as $s\to\infty$.

\section{Multisqueezed states under decoherence}\label{app:decoh_ms}

    \begin{figure}
        \centering
        \includegraphics[width=\linewidth]{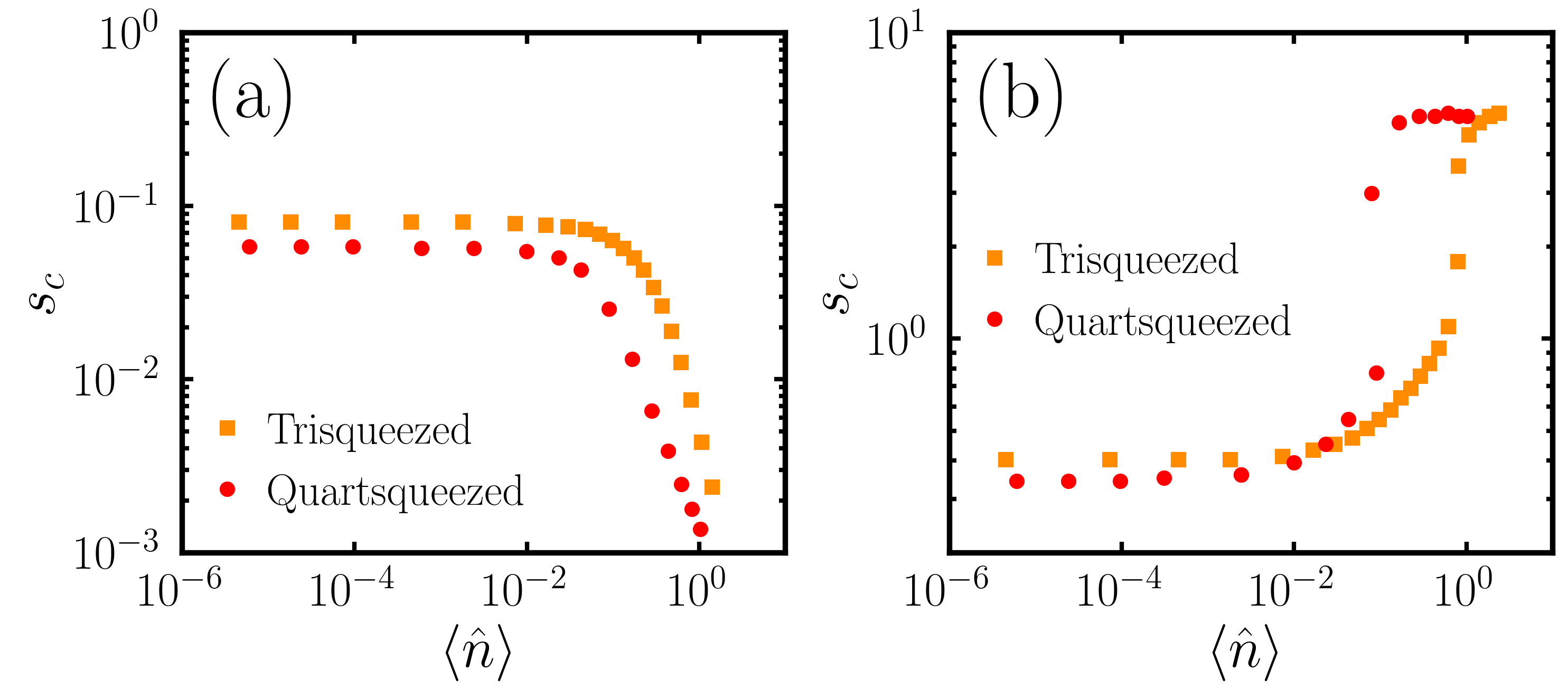}
        \caption{(a) Critical value of the pure dephasing noise strength $s_c$ above which multisqueezed states surpass the QFI of standard squeezed states, as a function of the occupation $\langle \hat{n}\rangle$. Orange squares and red circles correspond to trisqueezed and quartsqueezed states, respectively. Similarly, panel (b) shows the critical value of the zero-temperature noise strength $s_c$, with the same format as in (a).}
    \label{fig:decoherence_multisqueezed}
    \end{figure}

    As stated in the main text, the metrological performance of multisqueezed states under the influence of pure dephasing and zero-temperature damping noise is similar to \textit{m}th-phase states. Here we provide further details about their behavior, analyzing only the region of occupations at which there is an advantage in the noiseless case, that is, $F^{ms}_Q > F^S_Q$ at $s=0$ (cf. Fig.~\ref{fig:QFI_multisqueezed}). 

    For pure dephasing case, we compute the QFI of trisqueezed and quartsqueezed states under the channel $\Phi^{\rm pd}_s(\hat{\rho})$ as a function of the strength $s$, and for a fixed initial amplitude of the coherent state of the pump field, $\alpha=10$. Similar results can be obtained for different $\alpha$ values.  Then, we compare the results with the QFI of standard squeezed states under same conditions. As expected, dephasing has a stronger effect on multisqueezed than on standard squeezed states. This leads to a finite value of the noise strength $s_c$ up to which the QFI of these dephased non-Gaussian states is above dephased standard squeezed states. For $s>s_c$, dephased multisqueezed states lose their potential metrological advantage. In Fig.~\ref{fig:decoherence_multisqueezed}(a), we show $s_c$ as a function of the occupation $\langle \hn \rangle$ for trisqueezed and quartsqueezed states. In the SQL regime ($\langle \hn \rangle \lesssim 10^{-1}$ and $\langle \hn \rangle \lesssim 10^{-2}$ for trisqueezed and quartsqueezed states, respectively), the critical value $s_c$ remains approximately constant. 
    However, as for \textit{m}th-phase states (cf. Fig.~\ref{fig:pure-dephasing_mphase}(b)) beyond the SQL regime, multisqueezed states display a larger metrological advantage for $s=0$, but they also become increasingly fragile against dephasing, as shown by a decreasing $s_c$. 

    We proceed in a similar manner for  zero-temperature damping channel $\Phi^{{\rm zd}}_s(\hat{\rho})$. Again, we find the critical values of the noise strength $s_c$ above which the QFI of these non-Gaussian states falls below that of standard squeezed states, under same conditions. Figure~\ref{fig:decoherence_multisqueezed}(b) shows $s_c$ as a function of the initial occupation $\langle \hn \rangle$ for trisqueezed and quartsqueezed states with $\alpha = 10$. As for \textit{m}th-phase states (cf. Fig.~\ref{fig:zeroTdamping_mphase}(b)), $s_c$ remains approximately constant during the SQL regime. Beyond this regime, $s_c$ increases for both cases, until it saturates at large occupations. However, in this case, the saturation occurs earlier in occupation for quartsqueezed states, and both states saturate to a similar value, $s_c \approx 0.5$. These results indicate a robust metrological advantage of multisqueezed states to zero-temperature noise. 


%

\end{document}